\documentclass[sigconf]{acmart}

\usepackage{multicol}
\usepackage[frozencache=true,cachedir=.]{minted}
\usepackage{url}
\usepackage{caption}
\usepackage{subcaption}
\usepackage{tabularx}

\AtBeginDocument{%
  }

\setcopyright{acmlicensed}
\copyrightyear{2026}
\acmYear{2026}
\acmDOI{XXXXXXX.XXXXXXX}
\acmConference[ICPE '26]{17th ACM/SPEC International Conference on Performance Engineering}{May 04--08,
  2026}{Florence, Italy}
\acmISBN{978-1-4503-XXXX-X/2018/06}

\begin{document}

%%
%% The "title" command has an optional parameter,
%% allowing the author to define a "short title" to be used in page headers.
%% \title{Wasure: Bridging the Gap in WebAssembly Benchmarking}
\title{Wasure: A Modular Toolkit for Comprehensive WebAssembly Benchmarking}

%%
%% The "author" command and its associated commands are used to define
%% the authors and their affiliations.
%% Of note is the shared affiliation of the first two authors, and the
%% "authornote" and "authornotemark" commands
%% used to denote shared contribution to the research.
\author{Riccardo Carissimi}
\email{riccardo.carissimi1@studenti.unimi.it}
\orcid{0009-0008-5882-6370}
\affiliation{%
  \institution{Università degli Studi di Milano}
  \city{Milan}
  \country{Italy}
}

\author{Ben L. Titzer}
\email{btitzer@andrew.cmu.edu}
\orcid{0000-0002-9690-2089}
\affiliation{%
  \institution{Carnegie Mellon University}
  \city{Pittsburgh}
  \state{PA}
  \country{USA}
}

%%
%% By default, the full list of authors will be used in the page
%% headers. Often, this list is too long, and will overlap
%% other information printed in the page headers. This command allows
%% the author to define a more concise list
%% of authors' names for this purpose.
%%\renewcommand{\shortauthors}{}

%%
%% The abstract is a short summary of the work to be presented in the
%% article.
\begin{abstract}
 
%WebAssembly (Wasm) has become a key compilation target for portable and efficient execution across diverse platforms. However, systematically benchmarking its many engines remains difficult due to fragmentation and tooling limitations. This paper introduces \textit{Wasure}, a modular and extensible command-line toolkit that simplifies the execution and comparison of WebAssembly benchmarks across engines. To complement performance evaluation, we conducted a dynamic analysis of the benchmark suite included with Wasure.  Our analysis highlights substantial differences in code coverage, control flow, and execution patterns, underscoring the value of benchmark diversity. Wasure aims to support researchers and developers in performing more systematic, transparent, and insightful evaluations of WebAssembly engines.

WebAssembly (Wasm) has become a key compilation target for portable and efficient execution across diverse platforms. Benchmarking its performance, however, is a multi-dimensional challenge: it depends not only on the choice of runtime engines, but also on hardware architectures, application domains, source languages, benchmark suites, and runtime configurations. This paper introduces \textit{Wasure}, a modular and extensible command-line toolkit that simplifies the execution and comparison of WebAssembly benchmarks. To complement performance evaluation, we also conducted a dynamic analysis of the benchmark suites included with Wasure. Our analysis reveals substantial differences in code coverage, control flow, and execution patterns, emphasizing the need for benchmark diversity. Wasure aims to support researchers and developers in conducting more systematic, transparent, and insightful evaluations of WebAssembly engines.

% WebAssembly (Wasm) has become a key compilation target for portable and efficient execution across diverse platforms. However, systematically benchmarking its many engines remains difficult due to fragmentation and tooling limitations. In this paper, we present an in-depth dynamic analysis of a broad suite of WebAssembly benchmarks. Our analysis highlights substantial differences in code coverage, control flow, and execution patterns, highlighting the importance of benchmark diversity and the need for more structured evaluation tools. To this end, we developed \textit{Wasure}, a modular and extensible command-line toolkit that simplifies the execution and comparison of WebAssembly benchmarks across engines. Wasure aims to support researchers and developers in performing more systematic, transparent, and insightful evaluations of WebAssembly engines.

\end{abstract}

%%
%% The code below is generated by the tool at http://dl.acm.org/ccs.cfm.
%% Please copy and paste the code instead of the example below.
%%


%%
%% Keywords. The author(s) should pick words that accurately describe
%% the work being presented. Separate the keywords with commas.
%% A "teaser" image appears between the author and affiliation
%% information and the body of the document, and typically spans the
%% page.
%%\begin{teaserfigure}
%%  \includegraphics[width=\textwidth]{sampleteaser}
%%  \caption{Seattle Mariners at Spring Training, 2010.}
%%  \Description{Enjoying the baseball game from the third-base
%%  seats. Ichiro Suzuki preparing to bat.}
%%  \label{fig:teaser}
%%\end{teaserfigure}

%%\received{20 February 2007}
%%\received[revised]{12 March 2009}
%%\received[accepted]{5 June 2009}

%%
%% This command processes the author and affiliation and title
%% information and builds the first part of the formatted document.

%% To publish on Arxiv
%% https://elliothe.github.io/paper%20submission/2021/12/15/how_to_arxiv.html
\settopmatter{printacmref=false}
\setcopyright{none}
\renewcommand\footnotetextcopyrightpermission[1]{}
\pagestyle{plain}

\maketitle

\section{Introduction}

WebAssembly \cite{haas2017bringing} is a portable, low-level bytecode format that acts as a compilation target for various programming languages, including C/C++, Rust, AssemblyScript, Java, Scala, Kotlin, OCaml and others. Originally developed for the web, Wasm has expanded into broader domains such as cloud \cite{varda2018} and edge computing \cite{fastlywhitepaper, nieke2021}, the Internet of Things (IoT) \cite{li2021, liu2021}, and embedded or industrial systems \cite{nakazake2022}. It is increasingly being adopted as the primary sandboxing technology in emerging computing platforms due to its execution model, which enforces strong isolation between module instances. Designed for efficiency at both load and runtime, Wasm benefits from multiple high-performance implementations.

Wasm offers strong safety guarantees through a rigorously defined formal specification \cite{wasmspec2022}, a sound, mechanically-verified type system \cite{watt2018}, and verified implementations \cite{bosemiya2022}. Its runtime performance is near-native, achieved through ahead-of-time (AOT) or just-in-time (JIT) compilation. Unlike dynamic binary translation, Wasm simplifies the compilation process by structuring code into modules and functions rather than arbitrary machine code. While JIT compilers deliver high performance, some engines, such as JavaScriptCore and wasm3, opt for interpreters to minimize startup time or memory usage. Interpreters also enhance debuggability and introspection capabilities, as demonstrated by recent advancements in fast in-place interpreter designs like those in the Wizard Research Engine \cite{titzer2022} and JavaScriptCore.

\subsection{Wasm Development Workflow}

% Wasm binaries are typically produced by compiling high-level programming languages like C, C++, and Rust using frontend compilers. These compilers generate \texttt{.wasm} files, and in web contexts, they may also produce JavaScript "glue code" to facilitate integration with web APIs. In browsers, Wasm modules are executed within JavaScript engines, while in non-web environments, the binaries are run using standalone Wasm engines deployed directly on operating systems.

Wasm binaries are typically produced by compiling high-level programming languages like C, C++, and Rust using frontend compilers. These compilers generate \texttt{.wasm} files, which are a binary representation of a module's code and data.

Figure \ref{fig:wasm_example} shows a small code example in C, the corresponding code in WebAssembly’s \texttt{.wat} text format, and the same code in WebAssembly’s \texttt{.wasm} binary format.

The virtual machine or engine which runs Wasm may be browser-based, a standalone runtime environment, or a plugin in a larger application. This flexibility allows developers to reuse a single compilation artifact across different platforms in different scenarios.

\begin{figure}[h]
\centering
\begin{multicols}{3}
\begin{minipage}{0.3\textwidth}

\begin{minted}[fontsize=\scriptsize]{c}
int increment(int x)
{
    return x + 1;
}
\end{minted}

\textbf{(a) C Source Code.}

\end{minipage}

\begin{minipage}{0.3\textwidth}

\begin{minted}[fontsize=\scriptsize]{wat}
(func (param i32)
      (result i32)
  local.get 0
  i32.const 1
  i32.add
)
\end{minted}

\textbf{(b) WebAssembly\\ text format.}

\end{minipage}

\begin{minipage}{0.3\textwidth}

\begin{minted}[fontsize=\scriptsize]{c}
[...] // header with
      // type info
20 00 // local.get 0
41 01 // i32.const 1
6a    // i32.add
0b    // end
\end{minted}

\textbf{(c) Wasm\\ binary format.}

\end{minipage}
\end{multicols}

\caption{Example of a function compiled to WebAssembly.}
\label{fig:wasm_example}
\Description{Three columns in which the first one has a simple sum function written in C. The second one has the same function translated to WAT. The third one has the same function written in WASM bytecode.}
\end{figure}

\subsection{WebAssembly Engines}

Wasm engines are responsible for loading, validating and executing Wasm modules. Wasm bytecode runs on a stack-based virtual machine (VM) with an execution stack for activations of functions and one or more large byte-oriented memories for efficient data storage. The stack is not directly addressable, meaning errant memory accesses cannot corrupt control flow. This engine implements all operations that are defined by the Wasm binary specification, most of which map directly onto hardware instructions. The engine makes interactions between Wasm modules and their host environments frictionless \cite{mccallum2022}. Engines can either run as standalone applications or in a web browser, where it's embedded within the JavaScript engine and operates in conjunction with JavaScript code that give access to Web APIs.

Engines execute Wasm binaries either in isolation or as components embedded within host applications, often with capabilities for host function imports and system-level integration. Thanks to the availability of both browser and standalone engines, developers can build applications that are both portable and performant. This is a key factor in Wasm adoption. 

\subsubsection{Architecture}

The architecture of a Wasm engine includes four essential components: a parser (including validation), one or more execution tiers (such as an interpreter or compiler), a runtime environment, and an interface to the host system. Wasm code can be executed by interpreting the instructions or by compiling them to host (machine) code. Some engines — such as JavaScriptCore, wasm3 and Wizard — rely on interpretation. Others, like wasmtime \cite{wasmtime} and wasmer \cite{wasmer}, perform a two-phase compilation by compiling first to an intermediate representation (IR), then translate the IR into architecture-specific code.

Standalone engines, like wasmtime and wasmer, can also be embedded in applications developed in high-level programming languages. This makes developers able to run Wasm modules in various contexts. Some of these engines also include additional utilities, such as module manipulation tools, caching of AOT code, and more advanced developer tooling. 

\paragraph{WebAssembly Interpreter}
One way to execute Wasm binaries is through an interpreter, which processes each instruction in sequence directly from the bytecode. This avoids the need to translate the code into native machine instructions, resulting in faster startup times. However, because the code is interpreted rather than compiled, runtime performance is generally slower compared to executing native code.

\paragraph{WebAssembly Compiler}
Another approach is to compile the Wasm code to native machine instructions before or during the execution. Wasm compilers are usually categorized in two groups: just-in-time (JIT) and ahead-of-time (AOT) compilers. Each engine can have different backends to perform such operations. Tools like wasmer~\cite{wasmer} employ different backends, such as Singlepass, Cranelift or LLVM. The output of the compilation step differs based on the architecture and operating system, even if the input Wasm binary is the same. 

\subsection{System Level Interaction}
To enable system-level interaction, web-based Wasm engines invoke JavaScript bindings to access Web APIs. In other contexts, standalone engines provide different APIs, such as the WebAssembly System Interface (WASI)~\cite{wasi2022}. WASI abstracts common operating system functionalities, such as I/O and timers, in order to offer a uniform system interface across multiple platforms. It acts as a mediator between Wasm's sandbox and the host environment, granting safe access OS-level features. 

In other situations, purpose-built APIs are used.
To address limitations in system-level interactions, Ramesh et al. proposed WALI~\cite{ramesh2025} (WebAssembly Linux Interface). It virtualizes the bottom layer of userspace to directly expose OS syscalls to Wasm modules without compromising intra-process sandboxing.

\section{Approach}

We now present Wasure, our command-line toolkit designed to streamline WebAssembly benchmarking. We begin by outlining the design goals that guided its development (Section~\ref{sec:design-goals}), then introduce the high level architecture (Section~\ref{sec:system-architecture}). We proceed to describe its core components (Section~\ref{sec:core-modules}), introduce the runtimes included in the tool \ref{sec:selected-runtimes} and emphasize the extensibility of the project (Section~\ref{sec:extensibility}).

\subsection{Design Goals} \label{sec:design-goals}

The design of Wasure was guided by a set of foundational goals intended to make benchmarking WebAssembly engines both effective and accessible. Given the growing diversity of WebAssembly engines and the need for reproducible, cross-engine comparisons, Wasure aims to serve as a robust yet flexible tool for researchers and developers. Below we outline the primary design goals that informed the development of the toolkit.

\paragraph{Ease of Use}
A key goal of Wasure is to provide a frictionless user experience. This is done via an intuitive command-line interface that makes users able to run benchmarks with minimal effort. Subcommands and flags follow a consistent syntax. 

\paragraph{Extensibility}
Given the continued evolution of the WebAssembly ecosystem, Wasure was built with extensibility in mind. Both runtimes and benchmarks can be added through simple configuration changes. Users can also use several configurations of a runtime by leveraging \emph{subruntimes}.

\paragraph{Portability}
Wasure is implemented in Python and designed to run on Unix-like systems, namely Linux and macOS. While Windows support is currently limited, the modularity of the codebase leaves room for future cross-platform compatibility.

\paragraph{Accuracy}
Precise measurements are essential when evaluating WebAssembly runtimes. Wasure was developed to run benchmarks with minimal overhead, also supporting timeouts and multiple repetitions of the same benchmark. 

\subsection{System Architecture} \label{sec:system-architecture}

\begin{figure}[h]
  \centering
  \includegraphics[width=\linewidth]{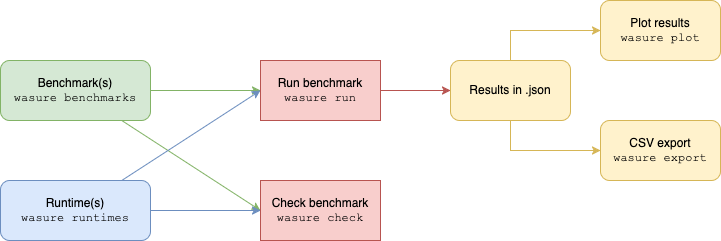}
  \caption{Wasure core modules' pipeline}
  \Description{Pipeline of wasure with all the core modules connected in three phases: preparation, execution and evaluation phase}
  \label{fig:wasure-pipeline}
\end{figure}

Wasure is designed as a modular command-line toolkit. As illustrated in Figure~\ref{fig:wasure-pipeline}, the system is organized into three distinct operational phases:

\begin{itemize}
\item \textit{Preparation Phase}: This phase provides utilities for setting up the benchmarking environment. Users can inspect and manage the default benchmark suite. They are also able to install or register WebAssembly runtimes and configure them. Runtime management supports both automatic installation and manual registration of engines already present on the system.
    
\item \textit{Execution Phase}: Benchmarks can be executed across multiple runtimes, either individually or in batch mode. During execution, Wasure keeps track of the time taken by the benchmark and records the peak values of both resident set size (RSS) and virtual memory size (VMS). Additionally, users can verify runtime compatibility with selected benchmarks using the \texttt{check} command. This is particularly useful for assessing support for WebAssembly features or WASI proposals. 

\item \textit{Evaluation Phase}: After execution, results are stored in structured JSON files. Wasure supports exporting these results to CSV for further analysis as well as rendering plots for visualization. This modular output format makes it easy to integrate with external data analysis workflows.
\end{itemize}

\subsection{Core Modules} \label{sec:core-modules}

The main entrypoint to Wasure is the CLI interface. This interface is structured around modular subcommands, each responsible for a specific functionality.

\subsubsection{Benchmark Management}
Wasure organizes benchmarks using an extensible structure, where each benchmark is grouped by suite (e.g. mibench, ostrich) into named collections called "benchmark groups". 

%The list of available benchmarks can be queried using:

%\begin{verbatim}
%    wasure benchmarks list
%\end{verbatim}

% Benchmarks are described as JSON objects stored in per-group \texttt{benchmarks.json} files. Each object supports several fields to define the behavior and expected output of the benchmark, as described in Section~\ref{sec:extensibility}.

\subsubsection{Runtimes Management}
The \texttt{runtimes} module provides a unified interface for managing WebAssembly engines used in benchmarking workflows. Through a consistent set of subcommands, users can install, update, list, inspect, and remove supported engines, or register their own.

% Users can query the list of engines that are installable using the \texttt{wasure runtimes available} command. It will check all the instructions found in the \texttt{installers/} folder. An engine can be installed by calling \texttt{wasure runtimes install}. Similar commands are also available to remove, update and query the installed version of engines. 

% Installed engines are validated first by running their respective version command and then by running a dummy payload (\texttt{dummy/dummy.wasm}). If the engine fails to execute it correctly, the installation is considered invalid and rolled back. If a subruntime fails, it is removed from the runtime.

\subsubsection{Benchmark Execution Engine}
Benchmark execution in Wasure is orchestrated by the \texttt{run} module, which launches performance experiments across WebAssembly runtimes. This module abstracts the complexity of runtime invocation, resource tracking and result aggregation behind user-friendly command-line interface.

The execution engine supports running both individual benchmarks and group of benchmarks, with options for configuration such as the number of repetitions, memory usage tracking, selective runtime execution, and output control. 

%A typical usage might look like the following:

%\begin{verbatim}
%wasure run -b helloworld -r wasmtime --repeat 3 --memory
%\end{verbatim}

%This command executes the \texttt{helloworld} benchmark three times using the Wasmtime engine while also collecting memory usage statistics during each run.

%The engine allows flexible specification of which runtimes and benchmarks to include in an experiment. Users can run:

%\begin{itemize}
%  \item all available runtimes and benchmarks with the keywords \texttt{-\--runtimes all} and \texttt{-\--benchmarks all},
%  \item a subset of runtimes (e.g., only wasmer or wasmtime),
%  \item or specific benchmark files or named sets of benchmarks like \texttt{coremark/coremark-1000}.
%\end{itemize}

Some benchmarks also collect internal timings and compute statistics such as an overall score.
Wasure also supports configurable per-benchmark score parsing by extracting numerical values from outputs using regular expressions. 
%Consider a benchmark that produces a performance score in its output, such as:

%\begin{verbatim}
%Benchmark completed: score=2314.6 iterations/sec
%\end{verbatim}

%The benchmark configuration might include a \texttt{score-parser} field like:

%\begin{verbatim}
%"score-parser": "score=(?P<score>\\d+(\\.\\d+)?)"
%\end{verbatim}

%This expression is parsed enabling the engine to extract and store the numerical score for later analysis.

\subsubsection{Results Structure}
Once benchmark execution completes, results are saved into JSON files.  The benchmarking framework also supports exporting results to CSV format via the command \texttt{wasure export}. This enables seamless integration with data analysis libraries for reporting, filtering, and statistical processing.

% Each file is timestamped (e.g., \texttt{2025-06-11\_14-23-45.json}) to prevent overwrites. The top-level structure is a dictionary keyed by engine names. Each iteration object contains:

%\begin{itemize}
%  \item \texttt{elapsed\_time\_ns}: integer \( \text{(nanoseconds)} \) — time taken by the benchmark invocation.
%  \item \texttt{score}: float — parsed performance metric, 0 if not applicable or parsing failed.
%  \item \texttt{return\_code}: integer — the process exit code; benchmarks with non-zero codes typically record zero time and score.
%  \item \texttt{output} (optional): string — raw stdout and stderr, included only if \texttt{-\--no-store-output} is \textit{not} set.
%  \item \texttt{stats} (optional): object — contains parsed statistics such as memory usage and runtime-specific metrics.
%\end{itemize}

\subsubsection{Results Visualization}

The \texttt{plot} module of \texttt{wasure} is responsible for producing comparative plots of benchmark results across different engines. It is designed to provide a fast and seamless way of understanding the results from the benchmarks. Its design aims to automatically adapt the visualization strategy based on the characteristics of the input data.

Depending on whether multiple runtimes are present per benchmark, the module selects between normalized and absolute visualization:

\begin{itemize}
    \item \textit{Normalized mode:} When a benchmark is executed by more than one runtime, values are normalized for fair comparison. Scores are normalized by the highest score (higher is better), while execution times are normalized by the lowest time (lower is better).
    \item \textit{Absolute mode:} When each benchmark is executed by a single runtime, absolute values are used, as there is no need for normalization.
\end{itemize}

\subsubsection{Engine Feature Tracking}

One of the essential challenges when benchmarking WebAssembly engines is determining which engines support which features. WebAssembly and its surrounding ecosystem are evolving rapidly. Not all engines are capable of running every benchmark and some of them depend on experimental or non-standard features. To assist with this task, Wasure provides the \texttt{check} module, a utility designed to report feature support across multiple engines.

At its core, the \texttt{check} command executes selected benchmarks across one or more engines and observes the result of each run, focusing specifically on whether a given benchmark completes successfully. In order to make feature detection easier, two specific sets of benchmarks are provided:

\begin{itemize}
    \item \texttt{wasm-features} contains a collection of payloads designed to test support for specific WebAssembly features~\cite{wasm-features}. Failure to execute a payload indicates that the corresponding proposal is not supported. Most of these payloads originate from Google Chrome’s WebAssembly feature detection project\footnote{\url{https://github.com/GoogleChromeLabs/wasm-feature-detect}}.
    \item \texttt{wasi-proposals} includes a set of payloads intended to test support for individual WASI proposals~\cite{wasi-proposals}. %As with WebAssembly features, failure to run a payload suggests that the related proposal is not supported.
\end{itemize}

Using these sets of custom built benchmarks, users can rely on Wasure's \texttt{check} module to automatically reveal compatibility gaps.

\subsection{Selected Engines} \label{sec:selected-runtimes}

Table~\ref{table:selected-runtimes} provides an overview of engines that can be installed effortlessly using Wasure. These engines can be broadly classified based on whether they operate within or outside a web browser environment.

When executed in web environments, WebAssembly modules typically rely on JavaScript integration and web platform APIs, while adhering to the web's security constraints. All major browsers offer native Wasm support via their underlying JavaScript engines. Wasure, and therefore our tests, make use of the CLI version of these engines. 

As stated in Section~\ref{sec:extensibility}, engines can be added manually by either providing an installer or by using a one that is already installed on the system. Installers are included in Wasure as a way to ease the installation process and to provide users a working setup with low effort. 
%However, we believe that a core strength of Wasure is to benchmark and compare custom engines.

\begin{table}
    \begin{tabular}{@{}lcccc@{}}
        \toprule
        \textbf{Wasm runtime} & \textbf{standalone} & \textbf{interpreter} & \textbf{JIT} & \textbf{AoT} \\
        \midrule
        V8 (Chromium) & $\circ$ & $\circ$ & $\bullet$ & $\circ$ \\
        SpiderMonkey (Firefox)& $\circ$ & $\circ$ & $\bullet$ & $\bullet$ \\
        JavaScriptCore (Safari)& $\circ$ & $\bullet$ & $\circ$ & $\circ$ \\
        wasmtime & $\bullet$ & $\circ$ & $\bullet$ & $\bullet$ \\
        Wasmer & $\bullet$ & $\circ$ & $\bullet$ & $\bullet$ \\
        WAMR & $\bullet$ & $\bullet$ & $\bullet$ & $\bullet$ \\
        WasmEdge & $\bullet$ & $\circ$ & $\circ$ & $\bullet$ \\
        wasm3 & $\bullet$ & $\bullet$ & $\circ$ & $\circ$ \\
        Wizard & $\bullet$ & $\bullet$ & $\bullet$ & $\circ$ \\
        wasmi & $\bullet$ & $\bullet$ & $\circ$ & $\circ$ \\
        wazero & $\bullet$ & $\bullet$ & $\bullet$ & $\circ$ \\
        \bottomrule
        \multicolumn{5}{l}{\footnotesize * $\bullet$ and $\circ$ refer to the positive and negative value respectively.}
    \end{tabular}
    \caption{Engines for which Wasure provides an installer}
    \label{table:selected-runtimes}
\end{table}

\subsection{Extensibility} \label{sec:extensibility}

\paragraph{Adding New Benchmarks}
Wasure is designed to make it easy to extend the benchmark suite with custom workloads. In most cases, running a benchmark requires nothing more than a WebAssembly binary. The system supports executing raw  \texttt{.wasm} files directly, allowing users to quickly test arbitrary modules. As such, for simple scenarios, such as measuring the performance of a single function or observing runtime compatibility, no additional setup is needed.

However, there are situations where defining a benchmark becomes beneficial. If the benchmark requires passing arguments, validating output correctness, or extracting a numerical score from the output, Wasure provides a structured way to define these behaviors. % This is done through a simple JSON-based schema.

%For instance, users might want to define validation patterns to check the correctness of the output of benchmarks. This can be configured using the \texttt{output-validator} field:

%\begin{minted}{json}
%{
%  "name": "dhrystone1B",
%  "path": "dhrystone1B.wasm",
%  "output-validator": "Dhrystones per Second:"
%}
%\end{minted}

%By providing similar fields, users can provide arguments or parse the score of benchmarks and make use of all the features that Wasure provides.

\paragraph{Adding New Runtimes}
Wasure allows users to easily integrate new WebAssembly runtimes into the benchmarking framework, whether for testing performance, feature support, or runtime-specific behaviors. Like benchmarks, runtimes are managed through a simple JSON-based configuration system that aims to minimize boilerplate while remaining highly customizable.

% Each runtime is represented as a JSON object within the \texttt{runtimes} array. At a minimum, a runtime definition must include a name, a description, and the commands necessary to run and query it. For example, the following defines a minimal Wasmtime installation:

%\begin{minted}{json}
%{
%    "name": "wasmtime",
%    "desc": "Wasmtime is a ...",
%    "version-command": "wasmtime -V",
%    "command": "wasmtime {payload}"
%}
%\end{minted}

%More complex runtimes can define further behaviors, including support for AOT compilation, entry point customization, subruntime variants, and runtime-specific statistics parsing.

Runtimes can also define subruntimes which are additional configurations of a runtime that reuse the same executable but change flags, execution modes, or output parsing rules. For example, WasmEdge can be benchmarked under its interpreter, JIT, and AOT modes by registering subruntimes like \texttt{wasmedge-int} or \texttt{wasmedge-aot}. Subruntimes are useful for comparing internal modes of the same runtime without duplicating installation details.

\section{Dynamic Benchmark Analysis}

This chapter presents the metrics obtained using dynamic instrumentation on the benchmarks included with Wasure. We begin by outlining the motivations (Section~\ref{sec:analysis-motivation}), then introduce the tooling and setup used (Section~\ref{sec:analysis-setup}). We proceed to present obtained results (Sections~\ref{sec:analysis-results} and \ref{sec:pca}) and explore some insights and implications that our results show (Section~\ref{sec:analysis-implications}).

\subsection{Motivation and Objectives}
\label{sec:analysis-motivation}

While WebAssembly has rapidly gained traction as a portable and efficient compilation target, most existing benchmark evaluations tend to focus narrowly on static analysis \cite{hilbig2021} or runtime execution time \cite{lehmann2019, herrera2018, khan2014}. These offer valuable insights, but they often overlook the behavioral complexity of real-world execution patterns. To the best of our knowledge, dynamic characteristics such as function call frequencies, code coverage, or the diversity of operations executed at runtime are underexplored in the current body of WebAssembly literature.

Given that Wasure already provides a curated and diverse set of benchmarks for WebAssembly engine evaluation, it presented an ideal testbed for extending the scope of performance analysis beyond purely static profiling. The inclusion of dynamic instrumentation in this context allows for a deeper understanding of how different binaries behave when executed under realistic conditions.

To enable this, we leveraged the dynamic analysis mechanisms of Wizard, a research-oriented engine that allows flexible, non-intrusive instrumentation of WebAssembly programs with minimal overhead \cite{titzer2024}. This choice was motivated by its extensibility and its suitability for gathering fine-grained execution data at runtime.

The primary objective of this section is twofold. First, we aim to characterize the runtime behavior of each benchmark in the Wasure suite, using a selected set of dynamic metrics that expose control flow, memory usage, and structural patterns. Second, we seek to uncover trends and differences across benchmarks that might not be apparent through static inspection alone. Such information is valuable for both engine developers and future benchmarking efforts.

Ultimately, the broader motivation is to enrich the current understanding of WebAssembly program behavior in practical scenarios. By doing so, we hope to contribute not only to the evaluation of runtime performance but also to the design of more representative and informative benchmarks for the WebAssembly ecosystem.

\subsection{Tooling and Setup}
\label{sec:analysis-setup}

%DONE: talk about Wastrumentation - mention in related work
%DONE: talk about Whamm - mention in related work
In planning our dynamic analysis, we came across two tools capable of doing so on WebAssembly payloads. Wasabi \cite{lehmann2019} is a general-purpose framework that is able to perform static rewriting on WASM code. This approach modifies the payload ahead of time by injecting instrumentation code, which then emits runtime events during execution. It was the first toolkit capable of performing dynamic analysis on WebAssembly. More recently, Titzer et al. \cite{titzer2024} describe a research-oriented WebAssembly engine, Wizard, that achieves dynamic instrumentation below the bytecode level by using engine mechanisms. This enables monitors written for Wizard to observe and instrument all paths of execution, including dynamically generated or rarely executed code paths. This runtime instrumentation also allows for adaptive behavior, where instrumentation can respond to observed behaviors in real-time.

Having built Wasure to be flexible and extensible, integrating the Wizard research engine is a matter of adding a configuration in the appropriate directory. We created a specific subruntime that enables the monitors we intended to use. These monitors utilize Wizard's efficient probe mechanism that allows extensible event handling at desired points of program execution. The set of metrics captured during analysis is detailed in Table~\ref{tab:metrics}.

All benchmarks and analyses were performed on a dedicated server equipped with an Intel(R) Xeon(R) Platinum 8168 processor and 376 GiB of memory, running Ubuntu 20.04.6 LTS. Some benchmark groups are not included in the analysis results, as we were unable to instrument them using the Wizard engine. Additionally, using Wasabi as a fallback was not feasible, as it does not support modules compiled with the wasm-gc proposal. The versions of each engine used are listed in Table~\ref{tab:runtime-versions}.

\begin{table}[h]
  \centering
  \footnotesize
  \caption{Runtime versions used during experimentation}
  \label{tab:runtime-versions}
  \begin{tabularx}{\linewidth}{lX}
    \toprule
    \textbf{Runtime} & \textbf{Version} \\
    \midrule
    Wizard Engine & 25$\beta$.1 x86-64-linux \\
    Wasmtime & 33.0.0 \\
    Wasmer & 4.3.0 \\
    WasmEdge & 0.14.1 \\
    WAMR (iwasm) & 2.3.1 \\
    Wazero & 1.9.0 \\
    V8 & 13.9.191 \\
    JavaScriptCore (jsc) & v296425 \\
    SpiderMonkey & JavaScript-C140.0 \\
    \bottomrule
  \end{tabularx}
\end{table}

\begin{table}
  \footnotesize
  \caption{Metrics considered during benchmark analysis}
  \label{tab:metrics}
  \begin{tabularx}{\linewidth}{lX}
    \toprule
    Name & Description\\
    \midrule
    \texttt{reach\_50} & Number of instructions required to account for 50\% of the total execution time \\
    \texttt{reach\_75} & Number of instructions required to account for 75\% of the total execution time \\
    \texttt{reach\_90} & Number of instructions required to account for 90\% of the total execution time \\
    \texttt{reach\_95} & Number of instructions required to account for 95\% of the total execution time \\
    \texttt{reach\_99} & Number of instructions required to account for 99\% of the total execution time \\
    \texttt{reach\_100} & Number of instructions required to account for 100\% of the total execution time  \\
    \texttt{instr\_cov} & Percentage of covered instructions \\
    \texttt{block\_cov} & Percentage of covered blocks \\
    \texttt{func\_cov} & Percentage of covered functions \\
    \texttt{exec\_funcs} & Number of executed functions \\
    \texttt{exec\_inst} & Number of executed instructions \\
    \texttt{total\_funcs} & Number of functions \\
    \texttt{g\_reads} & Number of reads to global variables \\
    \texttt{g\_writes} & Number of writes to global variables \\
    \texttt{int\_ops} & Number of type \texttt{int} operations executed \\
    \texttt{float\_ops} & Number of type \texttt{float} operations executed \\
    \texttt{ind\_call} &  Number of \texttt{call\_indirect} operations executed\\
    \texttt{writes} & Number of writes to memory \\
    \texttt{reads} & Number of reads to memory \\
    \texttt{in\_first} & Percentage of time spent in the most executed function \\
    \texttt{total\_cycles} & Total number of cycle loops executed \\
    \texttt{time\_ns} & Mean time taken to execute across all runtimes \\
    \texttt{rss} & Mean resident set size (RSS) across all runtimes \\
    \texttt{vms} & Mean virtual memory size (VMS) across all runtimes \\
    \bottomrule
  \end{tabularx}
\end{table}

\subsection{Results and Discussion}
\label{sec:analysis-results}

%DONE use different colors and not shades of the same one for a group
\begin{figure}[h]
    \centering
    \includegraphics[width=\linewidth]{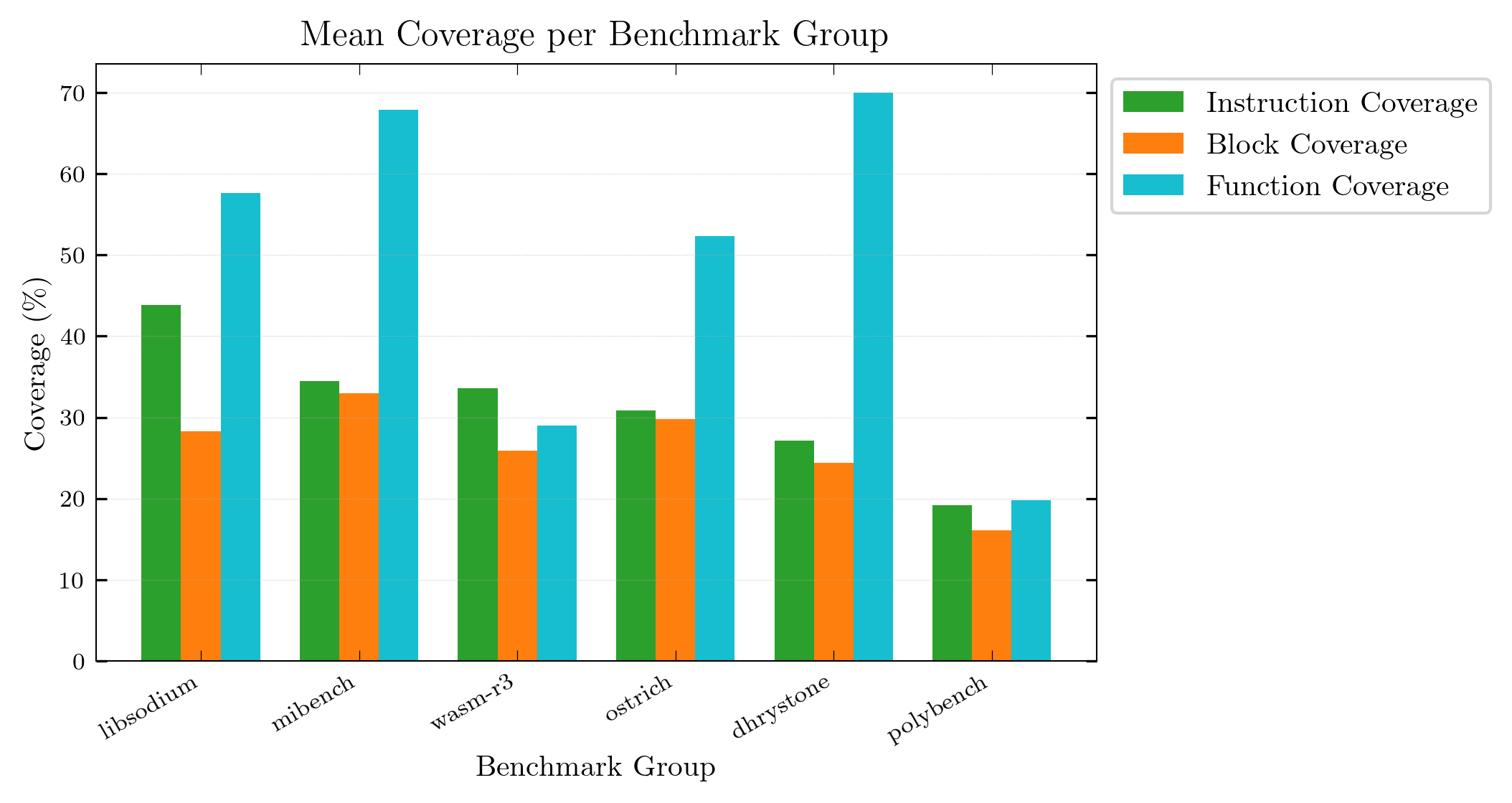}
    \caption{Mean dynamic instruction, block and function coverage across benchmark groups.}
    \Description{Grouped bar plot of the mean code, block and function coverage aross benchmark groups.}
    \label{fig:results-coverage}
\end{figure}

\paragraph{Coverage}
Figure~\ref{fig:results-coverage} shows the average dynamic instruction, block, and function coverage for each benchmark group. These metrics reflect the proportion of executed instructions, basic blocks, and functions relative to the total number that exist in each binary.

%DONE: analyze: https://link.springer.com/article/10.1007/s11390-010-9384-3
%DONE: analyze: https://www.sable.mcgill.ca/publications/techreports/sable-tr-2002-11.pdf
%DONE: cite: https://dl.acm.org/doi/10.1145/3030207.3030217
Overall, coverage is generally low across all benchmark groups, which is expected given that dynamic analysis only captures code paths actually executed during the test workloads. Among the benchmark groups, \texttt{mibench} and \texttt{dhrystone} exhibit the highest function coverage, exceeding 65\% and 70\% respectively. This suggests that their workloads exercise a larger portion of the available control flow, likely due to more comprehensive or diverse inputs.

%TODO: can we normalize the bar to be percentages and then add a data label for total number of functions?
\begin{figure}[h]
    \centering
    \includegraphics[width=\linewidth]{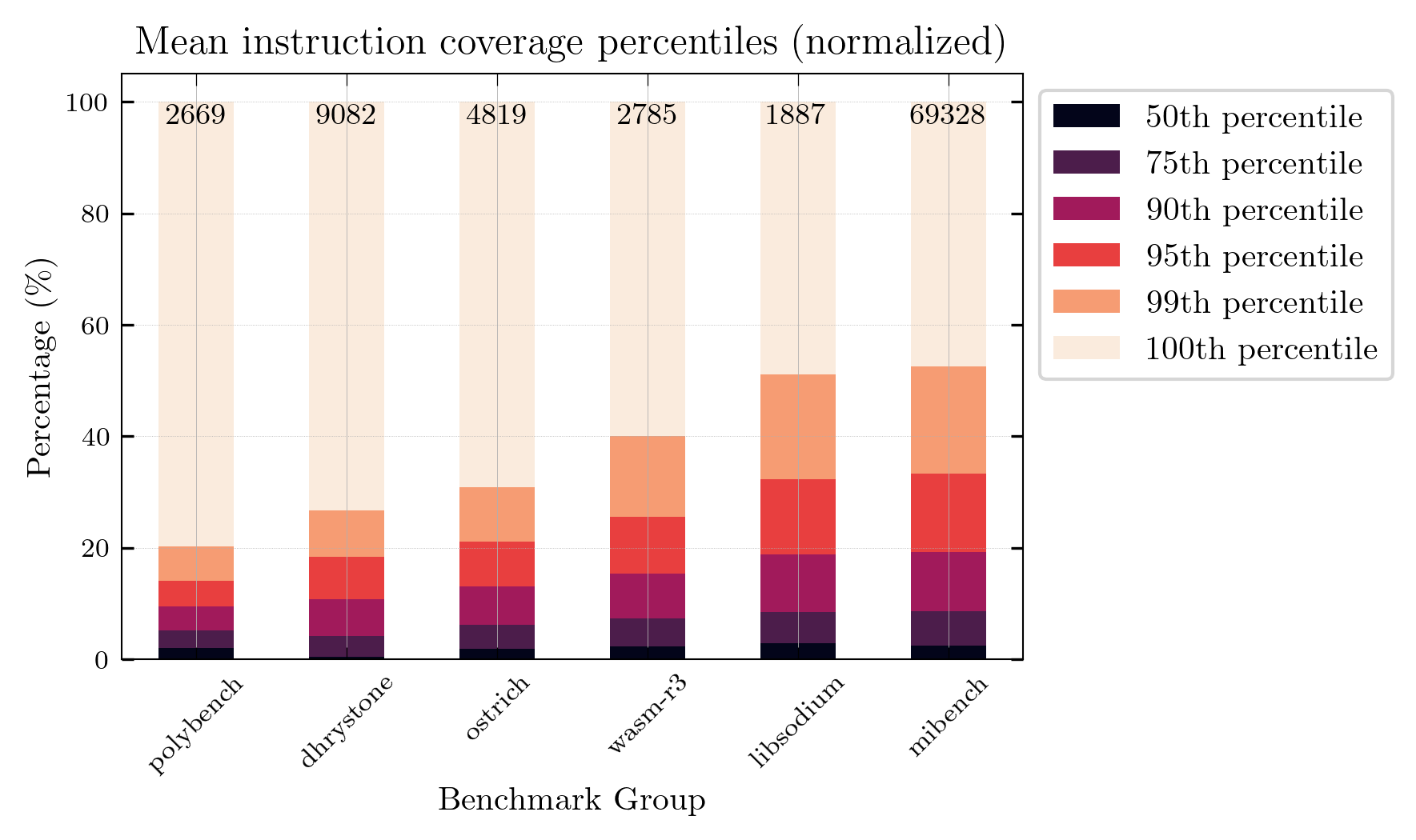}
    \caption{Normalized instruction hotness percentiles per benchmark. Each bar indicates the mean number of instructions required to account for 50\%, 75\%, 90\%, 95\%, and 100\% of total execution time per benchmark group.}
    \Description{Normalized stacked bar plot of the instruction hotness percentiles per benchmark group. wasm-r3 shows results that are an order of magnitude higher than the other benchmark groups.}
    \label{fig:results-function-percentiles}
\end{figure}

%DONE: more data points possible? no, its related to percentiles.
\begin{figure}[h]
    \centering
    \includegraphics[width=\linewidth]{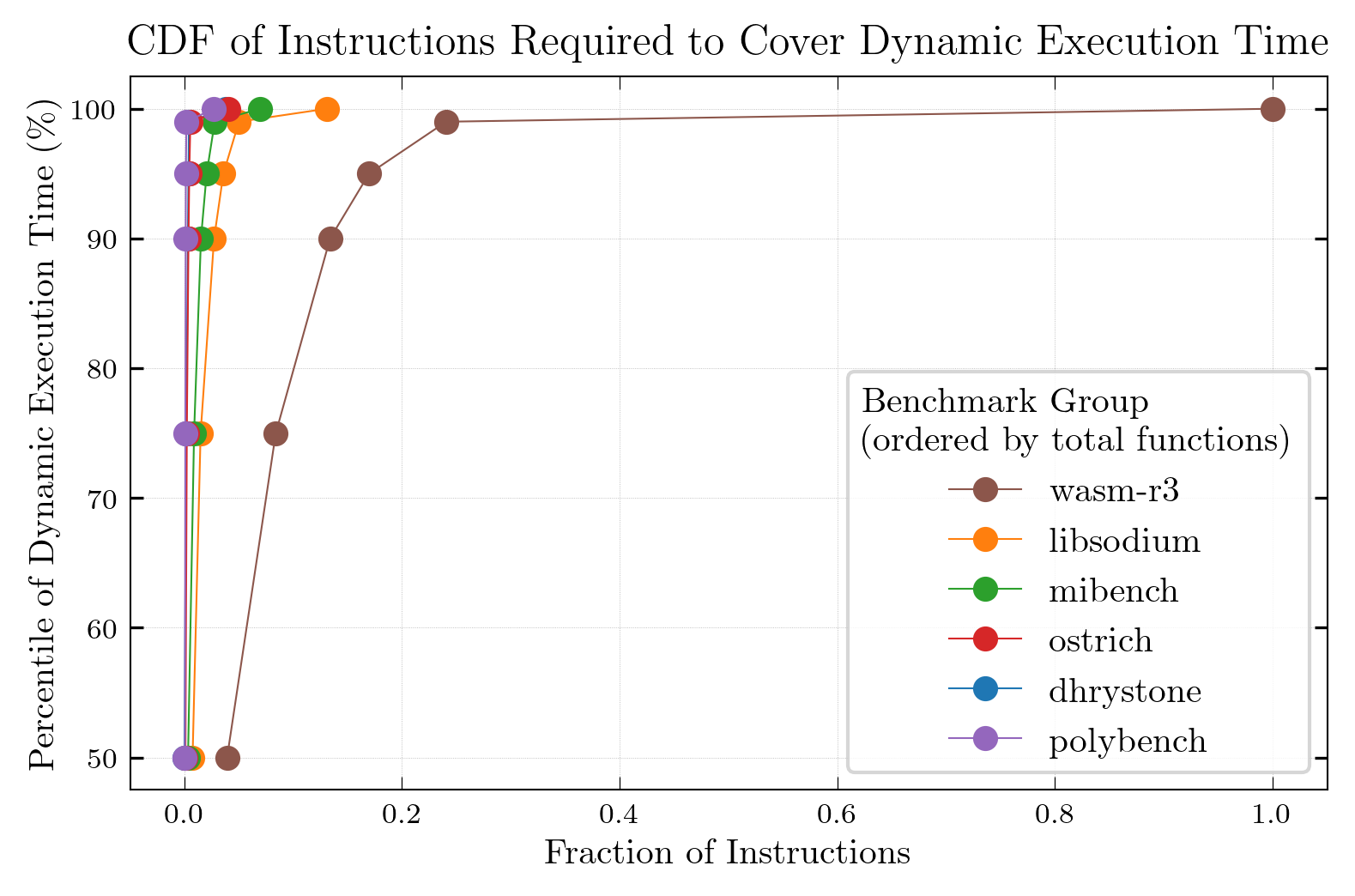}
    \caption{CDF of dynamic execution time covered by functions. The x-axis shows the fraction of total executed functions required to account for increasing percentages of dynamic execution time.}
    \Description{A cumulative distribution function plot showing how few functions contribute to most of the execution time in various benchmark groups. Most benchmarks reach over 95\% coverage with less than 10\% of functions, except wasm-r3.}
    \label{fig:results-function-percentiles-cdf}
\end{figure}

%DONE: instruction hotness or function? pick one. seems instructions.
% wizeng —mode=jit —metrics <program.wasm>
\paragraph{Instruction hotness.} Figure~\ref{fig:results-function-percentiles} illustrates the distribution of instruction hotness across benchmark groups by showing the normalized mean number of instructions needed to account for specific percentiles of the total execution time. Each benchmark group also shows the mean number of instructions to account for 100\% coverage of code. The results indicate that a relatively small subset of instructions is responsible for the majority of engine activity in all benchmarks.

The \texttt{wasm-r3} group stands out with an exceptionally high instruction count across all percentiles. It's an order of magnitude higher than any other benchmark group. This is a direct consequence of the nature of the \texttt{wasm-r3} suite: these benchmarks are automatically generated replayable payloads created using the Wasm-R3 tool. Wasm-R3 captures real-world web-based Wasm application executions, faithfully reproducing them in standalone form through record-and-replay instrumentation. Because these benchmarks include complete traces of original program logic along with emulated host interactions, they result in dense and expansive function settings.

In contrast, benchmarks like \texttt{dhrystone}, \texttt{ostrich}, and \texttt{polybench} show a much more compact distribution. A small number of instructions (often fewer than a few hundred) dominate their execution time, particularly at the 90th and 95th percentiles. This indicates the presence of distinct hot paths—functions that are heavily reused and critical to overall performance. Such behavior is typical in computation-heavy benchmarks where a few tight loops or kernels dominate runtime. 

%DONE: data does not make sense. try to rerun if have time, otherwise remove.
%
%\begin{figure}[h]
%    \centering
%    \includegraphics[width=\linewidth]{images/total-cycles-violin-ordered.png}
%    \caption{Distribution of total execution cycles across benchmark groups, ordered by median.}
%    \Description{Violin plot showing the distribution of total execution cycles per benchmark group. Each violin reflects the distribution shape and spread, with wider sections representing denser regions of the data. Groups are ordered left to right by increasing median cycles.}
%    \label{fig:results-cycles}
%\end{figure}

%\paragraph{Number of cycles.}
%Figure~\ref{fig:results-cycles} presents a violin plot showing the distribution of total execution cycles across benchmark groups.As expected, \texttt{polybench} reports the highest mean cycle count, reflecting its focus on large numerical kernels and tight loop computations. The distribution is relatively tight, reflecting the uniform nature of the workloads in this suite. By contrast, \texttt{mibench} displays a much broader distribution, indicating significant variability in runtime behavior across its constituent programs. This is expected given the suite’s heterogeneity, which spans domains such as automotive, networking, and consumer applications. \texttt{dhrystone} was excluded from this plot due to having only a single data point, which makes it unsuitable for distribution-based visualizations.

\begin{figure}[h]
    \centering
    \includegraphics[width=\linewidth]{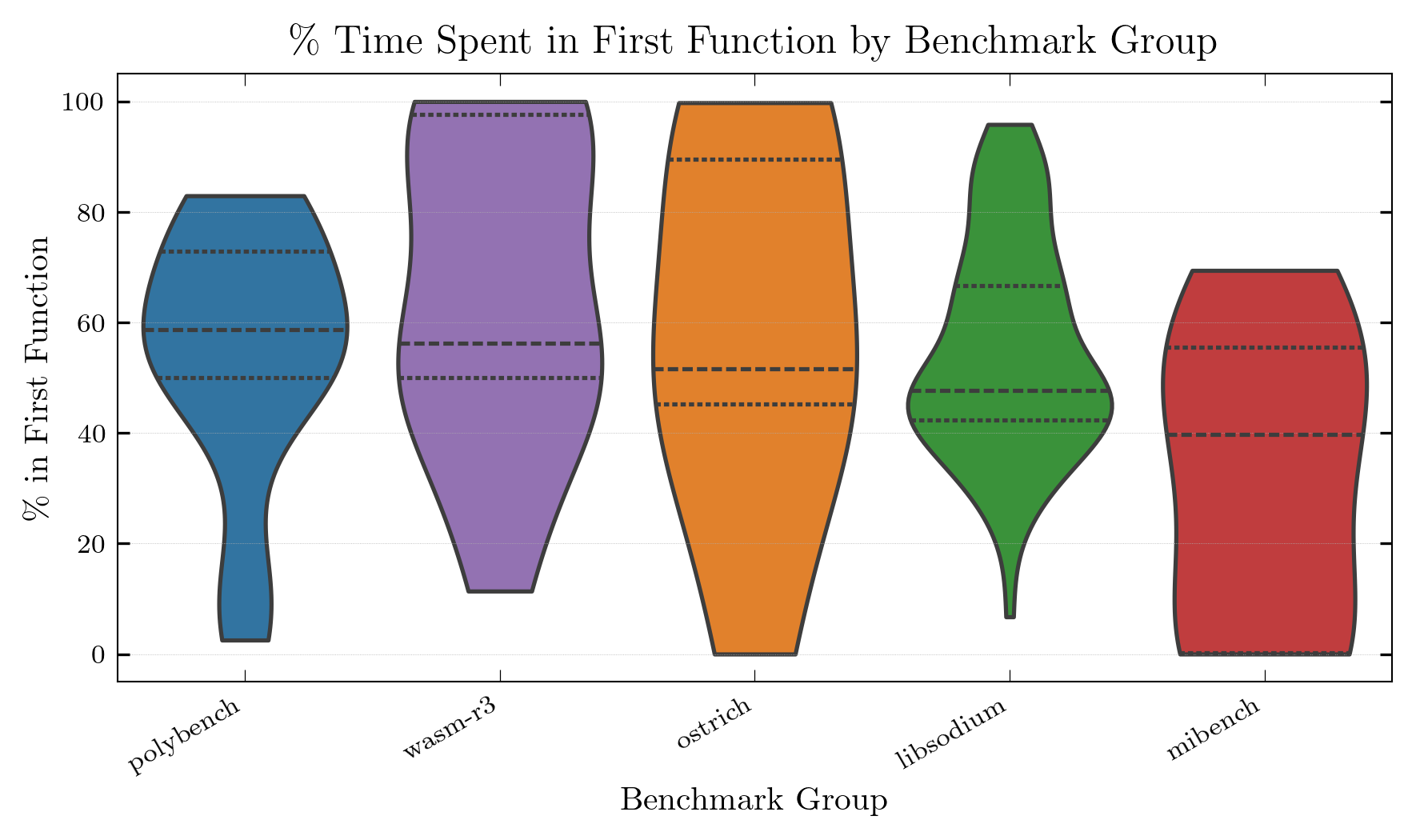}
    \caption{Distribution of percentage of total execution time spent in the most executed function, grouped by benchmark suite.}
    \Description{Violin plot showing the distribution of percentage time spent in the most executed function per benchmark group. Each violin represents the distribution shape and spread, with dashed lines indicating quartiles.}
    \label{fig:results-time-spent-one-function}
\end{figure}

\paragraph{Time spent in most executed function.}
Figure~\ref{fig:results-time-spent-one-function} shows the distribution of the percentage of total execution time spent in the most frequently executed function across benchmark groups. The \texttt{wasm-r3} and \texttt{ostrich} groups show the widest distributions, indicating a high degree of variability in execution profiles. In these cases, some benchmarks spend nearly all of their time in a single function, while others distribute time more evenly. This reflects the diverse nature of these suites: \texttt{wasm-r3} benchmarks are automatically generated and may include large serialized host interactions, while \texttt{ostrich} covers a mix of real-world program traces.

\begin{figure}[h]
    \centering
    \includegraphics[width=\linewidth]{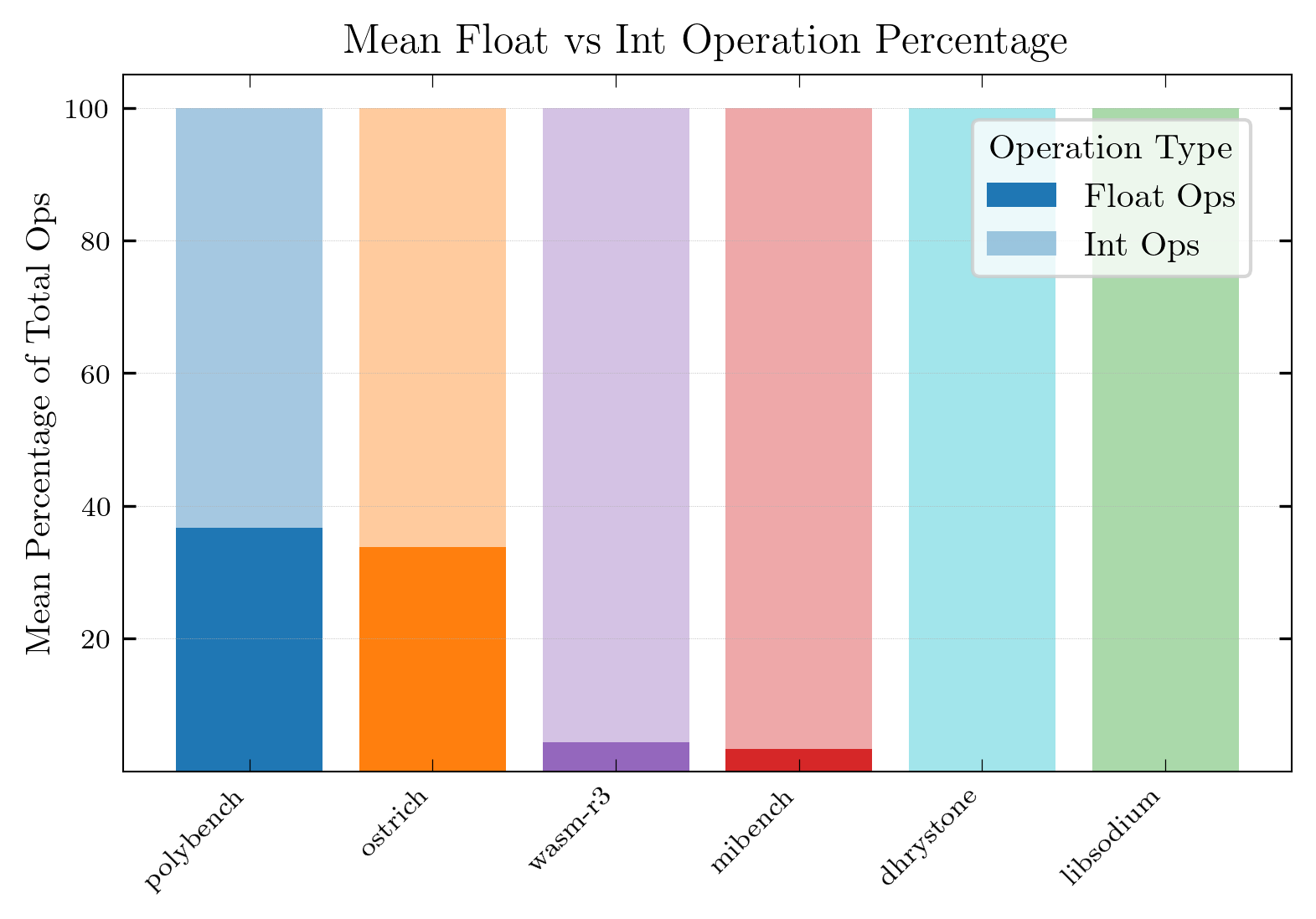}
    \caption{Mean ratio of floating-point to integer operations per benchmark group.}
    \Description{Bar chart that displays the mean proportion of floating-point versus integer operations across benchmark groups.}
    \label{fig:results-float-ops}
\end{figure}

\paragraph{Mean floating point operations.}
Figure~\ref{fig:results-float-ops} displays the mean proportion of floating-point versus integer operations across benchmark groups. As expected, \texttt{polybench} and \texttt{ostrich} exhibit the highest ratio of floating-point operations, reflecting their emphasis on scientific computing and numerical workloads. In contrast, \texttt{wasm-r3}, \texttt{mibench}, \texttt{dhrystone}, and \texttt{libsodium} are overwhelmingly dominated by integer operations. This dichotomy highlights the importance of selecting benchmarks based on application domain: while floating-point heavy benchmarks are useful for evaluating numerical kernels and vectorization strategies, integer-dominated workloads better represent real-world systems code and control logic.

\begin{figure}[h]
    \centering
    \includegraphics[width=\linewidth]{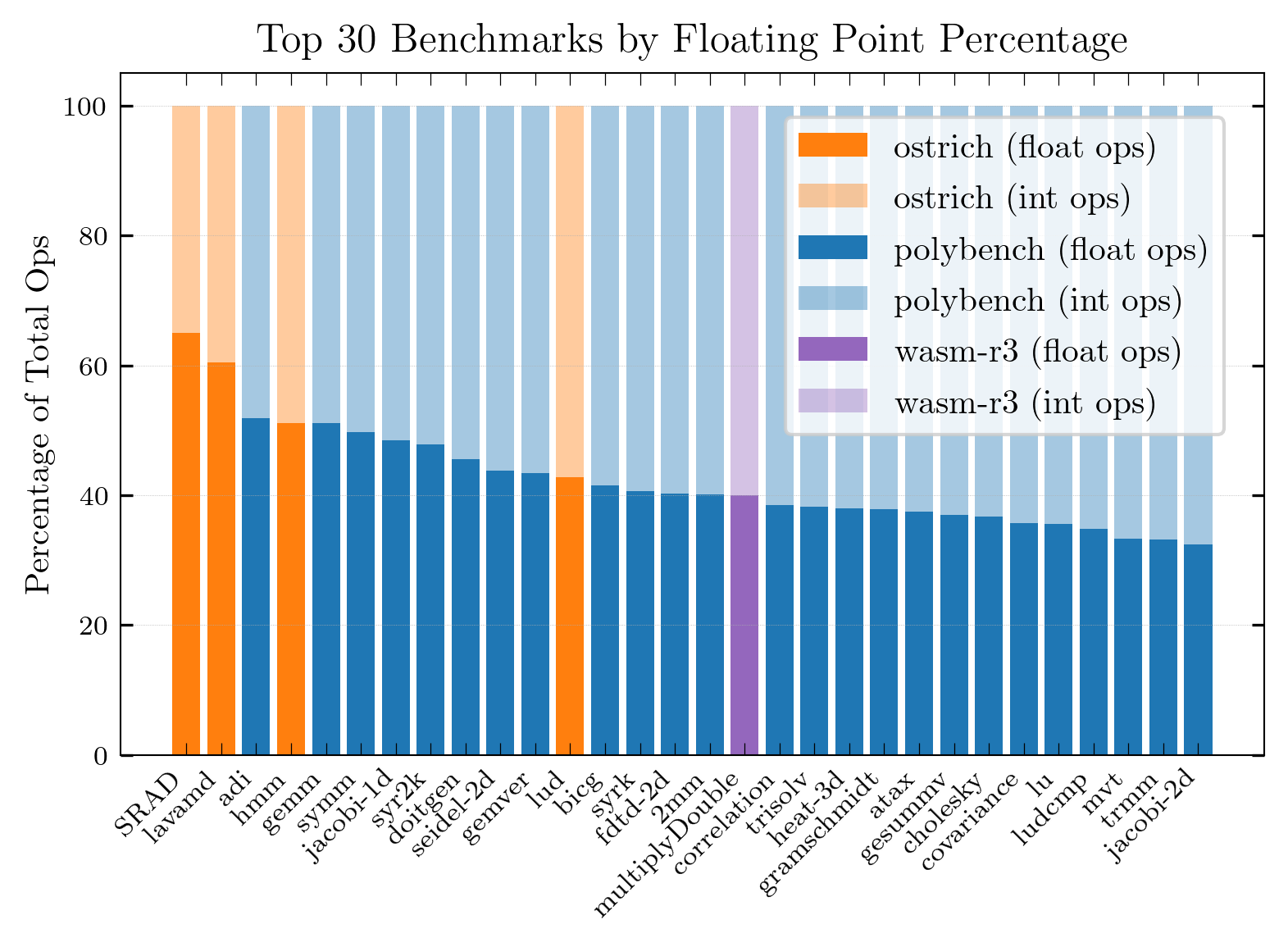}
    \caption{Top 30 benchmarks with the highest percentage of floating-point operations.}
    \Description{Bar chart that shows the top 30 individual benchmarks ranked by their percentage of floating-point instructions.}
    \label{fig:results-float-ops-top-30}
\end{figure}

\paragraph{Floating-point intensity by benchmark.}
Figure~\ref{fig:results-float-ops-top-30} breaks down the top 30 individual benchmarks ranked by their percentage of floating-point instructions. The results show clear group-level trends: \texttt{ostrich} benchmarks dominate the top of the ranking, with \texttt{SRAD} and \texttt{lavaMD} exceeding 60\% of their operations as floating-point. The lack of benchmarks coming from most of the groups analyzed highlights a sharp divide between numerical and systems-oriented workloads, emphasizing the importance of benchmark diversity when evaluating compiler or hardware optimizations targeting arithmetic-heavy execution.

%%\begin{figure}[h]
%%    \centering
%%    \includegraphics[width=\linewidth]{images/call-indirect-top-30.png}
%%    \caption{Top 30 benchmarks by number of indirect function calls (\texttt{CALL\_INDIRECT}).}
%%    \Description{Bar chart that shows the top 30 benchmarks with the highest dynamic count of call\_indirect instructions, plotted on a logarithmic scale.}
%%    \label{fig:results-call-indirect-top-30}
%%\end{figure}

\begin{figure}[h]
    \centering
    \includegraphics[width=\linewidth]{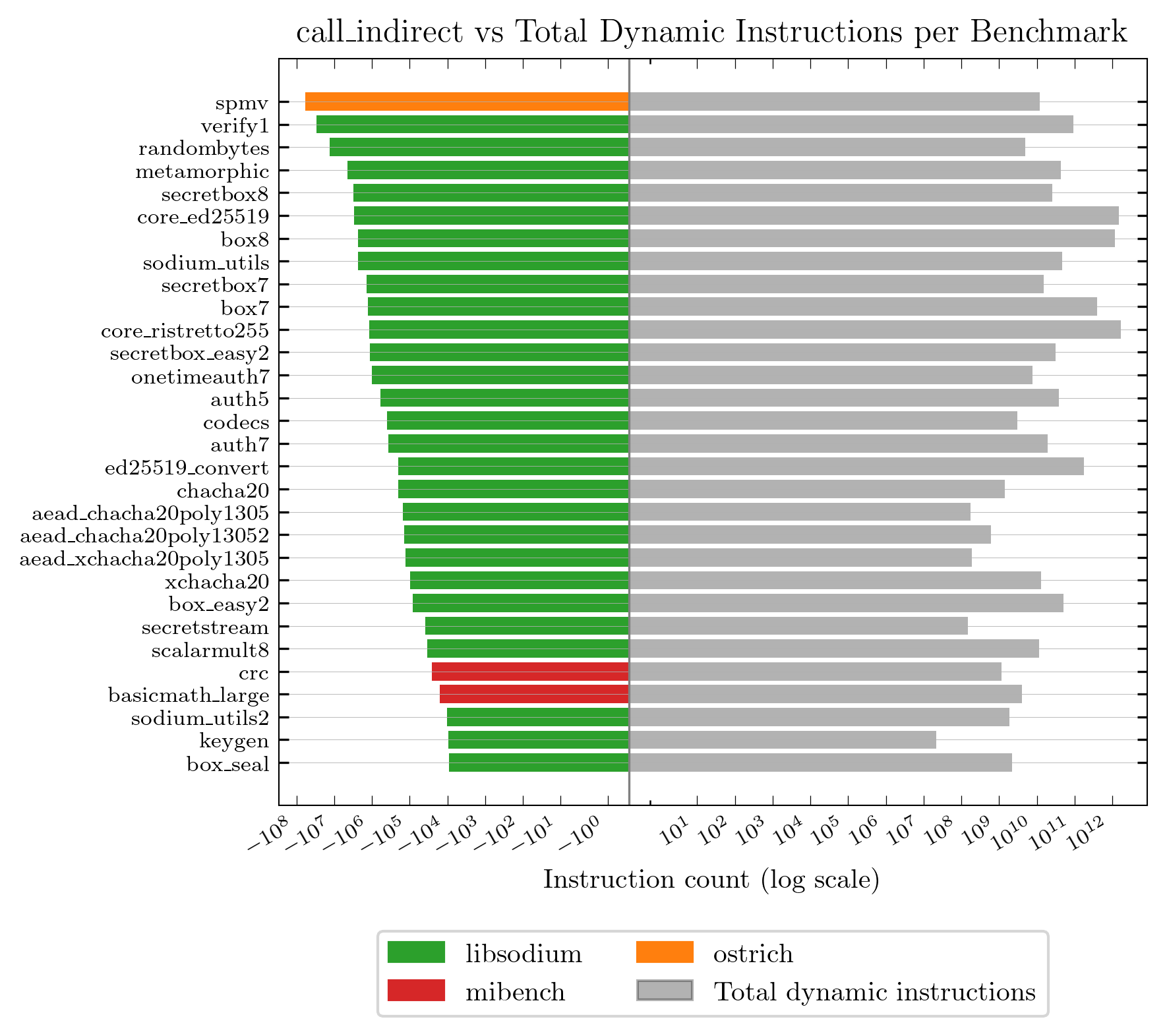}
    \caption{Top 30 benchmarks by number of indirect function calls (\texttt{CALL\_INDIRECT}) and their executed instructions count}
    \Description{Two bar charts that shows the top 30 benchmarks with the highest dynamic count of call\_indirect instructions on the left and their total number of dynamically executed instructions on the right, plotted on a logarithmic scale}
    \label{fig:results-call-indirect-vs-instructions}
\end{figure}

\paragraph{Indirect function call intensity.}
%%Figure~\ref{fig:results-call-indirect-top-30} shows the top 30 benchmarks with the highest dynamic count of \texttt{call\_indirect} instructions, plotted on a logarithmic scale. The figure reveals a striking disparity in indirect call usage across benchmarks. Notably, the \texttt{ostrich} benchmark \texttt{spmv} dominates the ranking by a wide margin, followed by \texttt{verify1} and \texttt{randombytes}, all of which show dynamic counts exceeding tens of millions. These benchmarks reflect application domains or code generation styles that rely heavily on indirect dispatch, such as dynamic kernels or cryptographic libraries.

Figure~\ref{fig:results-call-indirect-vs-instructions} shows the top 30 benchmarks with the highest dynamic count of \texttt{call\_indirect} instructions alongside their number of executed instructions, plotted on a logarithmic scale. The figure reveals a striking disparity in indirect call usage across benchmarks. Notably, the \texttt{ostrich} benchmark \texttt{spmv} dominates the ranking by a wide margin, followed by \texttt{verify1} and \texttt{randombytes}, all of which show dynamic counts exceeding tens of millions. These benchmarks reflect application domains or code generation styles that rely heavily on indirect dispatch, such as dynamic kernels or cryptographic libraries.

\begin{figure}[h]
    \centering
    \includegraphics[width=\linewidth]{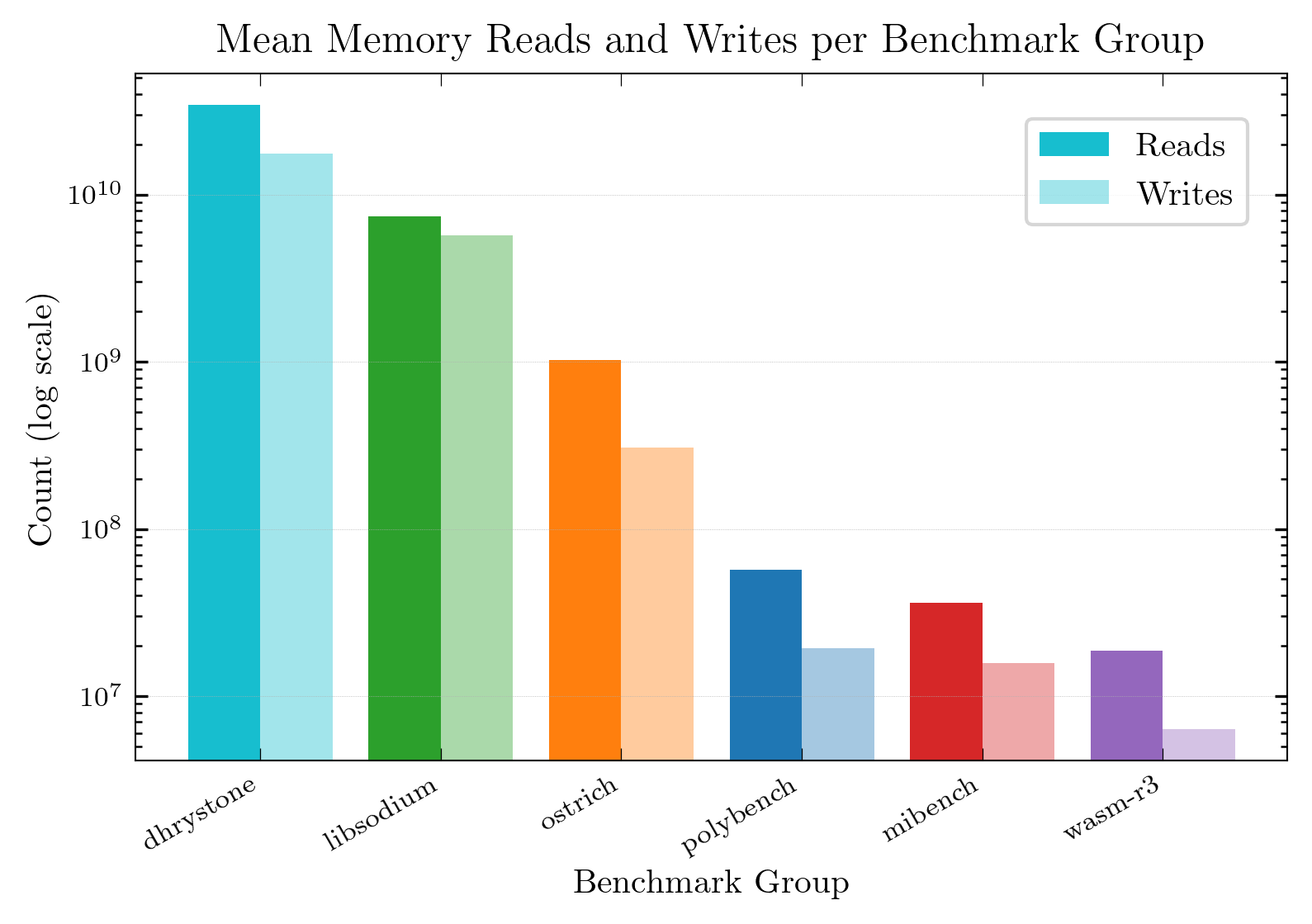}
    \caption{Total mean memory reads and writes aggregated per benchmark group.}
    \Description{Grouped bar chart displays the total mean number of memory reads and writes for each benchmark group, using a logarithmic scale.}
    \label{fig:results-memory}
\end{figure}

\paragraph{Memory access behavior.}
Figure~\ref{fig:results-memory} displays the total mean number of memory reads and writes for each benchmark group, using a logarithmic scale to accommodate the wide range of values. The most memory-intensive benchmarks are clearly \texttt{dhrystone} and \texttt{libsodium}, both of which exhibit read and write counts exceeding \(10^{10}\) and \(10^9\) respectively. The results suggest that while some benchmarks derive complexity from computational intensity and control flow others impose substantial demands on the memory subsystem, underscoring the need to analyze both aspects when evaluating WebAssembly performance profiles.

\begin{figure}[h]
    \centering
    \includegraphics[width=\linewidth]{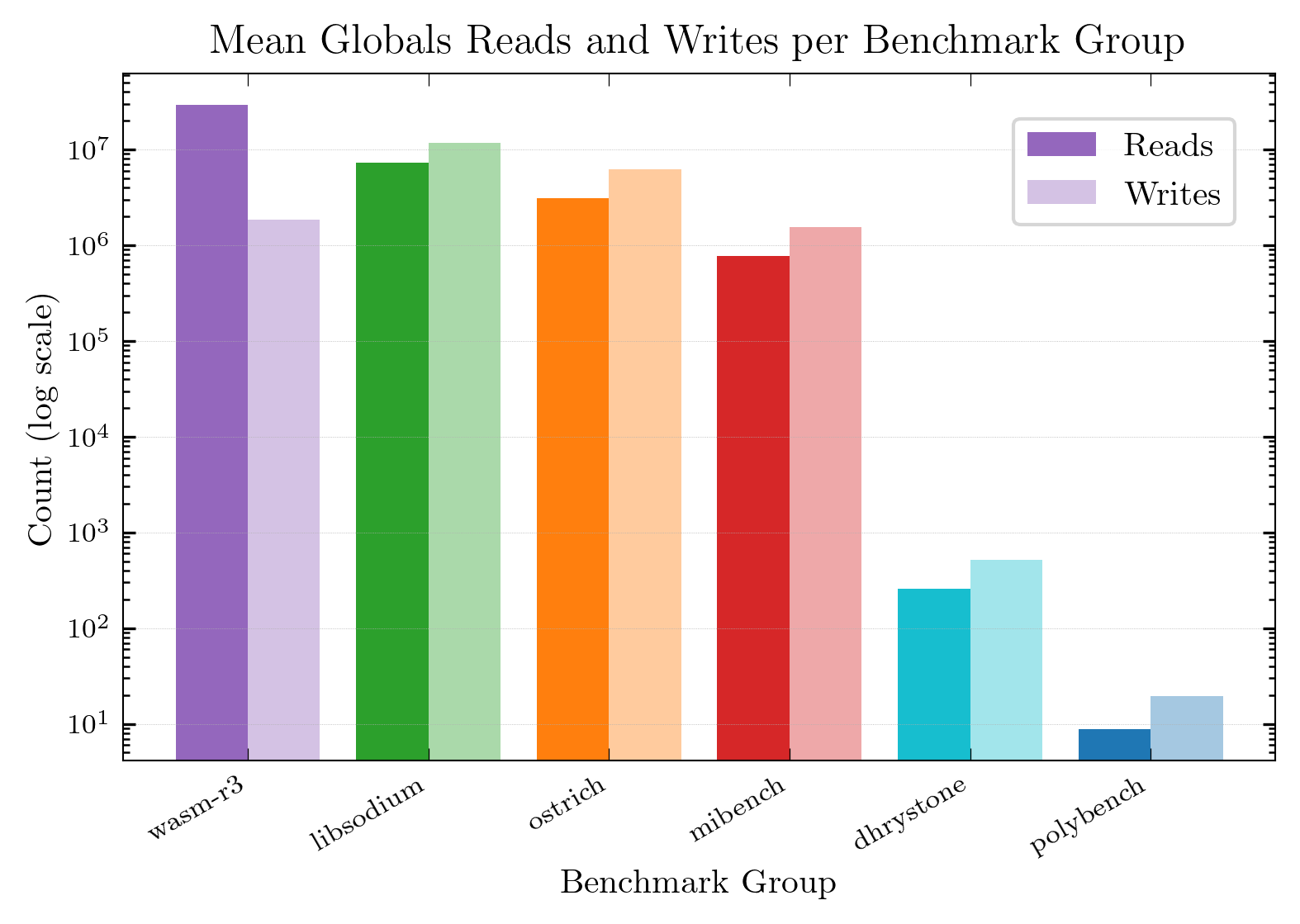}
    \caption{Mean number of reads and writes to global variables during execution across benchmark groups. }
    \Description{Grouped bar chart that presents the average number of reads and writes to global variables across the benchmark groups, shown on a logarithmic scale.}
    \label{fig:results-globals}
\end{figure}

\paragraph{Global variable access.}
Figure~\ref{fig:results-globals} presents the average number of reads and writes to global variables across the benchmark groups, shown on a logarithmic scale. The results highlight significant disparities in global variable usage patterns. The \texttt{wasm-r3} group dominates in terms of read operations, with mean read counts surpassing \(10^8\), while writes are noticeably fewer but still substantial. This high read activity aligns with its nature as a replayed trace of real-world applications, which tend to access a wide range of global state during execution.

\begin{figure}[h]
    \centering
    \includegraphics[width=\linewidth]{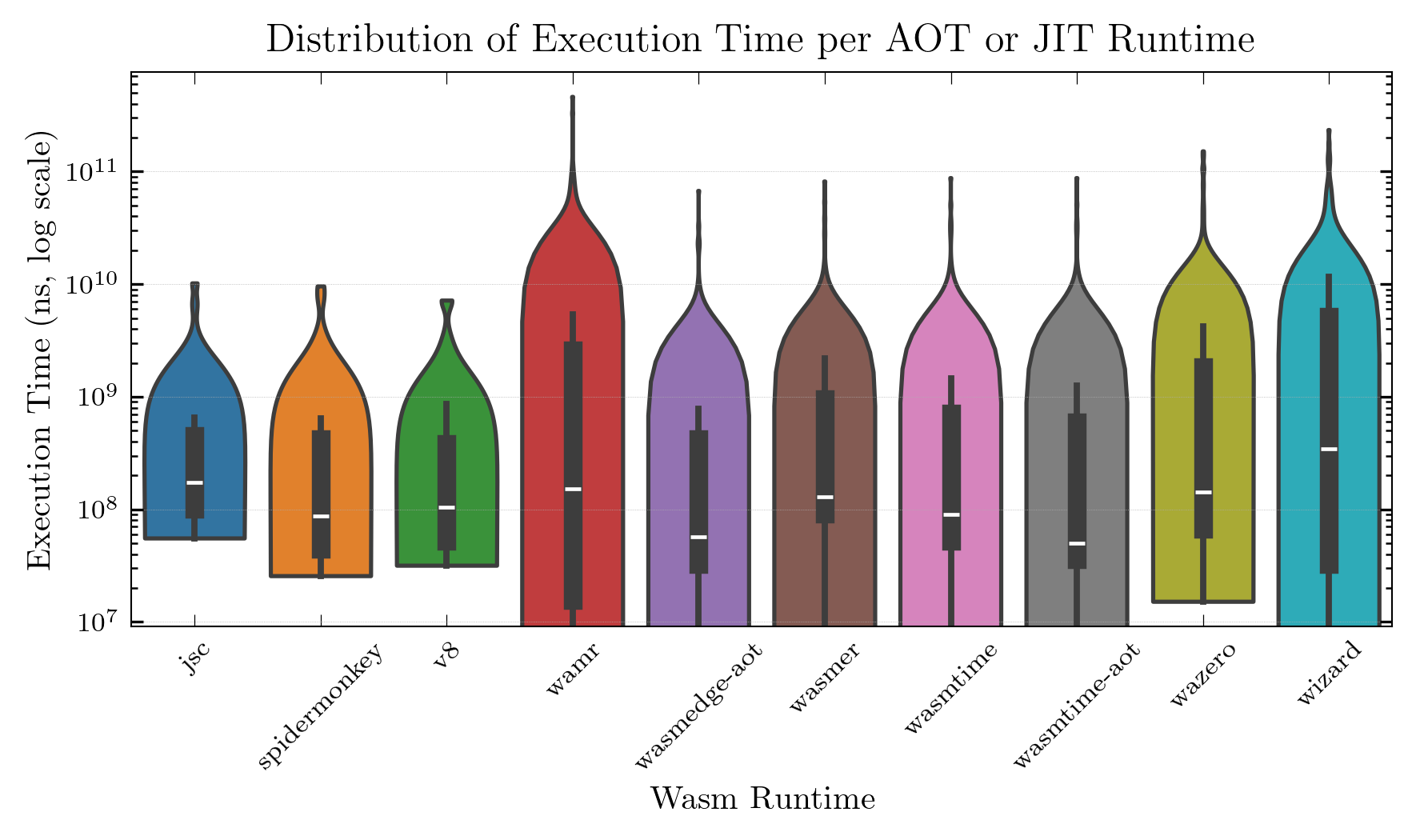}
    \caption{Distribution of execution time across WebAssembly engines.}
    \Description{Violin plot displaying execution time distributions per WebAssembly engine on a logarithmic scale. Includes engines such as jsc, spidermonkey, v8, wamr, wasmedge-aot, wasmtime, wazero, and wizard.}
    \label{fig:results-execution-time}
\end{figure}

\paragraph{Execution time distribution.}
Figure~\ref{fig:results-execution-time} presents the distribution of execution times across all WebAssembly engines, with values shown on a logarithmic scale. The violin plots capture both the central tendency and spread of execution durations observed in the benchmark suite. While some engines such as \texttt{jsc}, \texttt{spidermonkey}, and \texttt{v8} exhibit tighter, lower-centered distributions, others like \texttt{wamr}, \texttt{wizard}, and \texttt{wazero} display broader variability and heavier tails. The vertical range of each violin reflects the degree of performance fluctuation across benchmarks, with median execution times differing by up to two orders of magnitude between engines. These results emphasize that engine performance is not only dependent on average speed but also on consistency across workloads.

\begin{figure}[h]
    \centering
    \includegraphics[width=\linewidth]{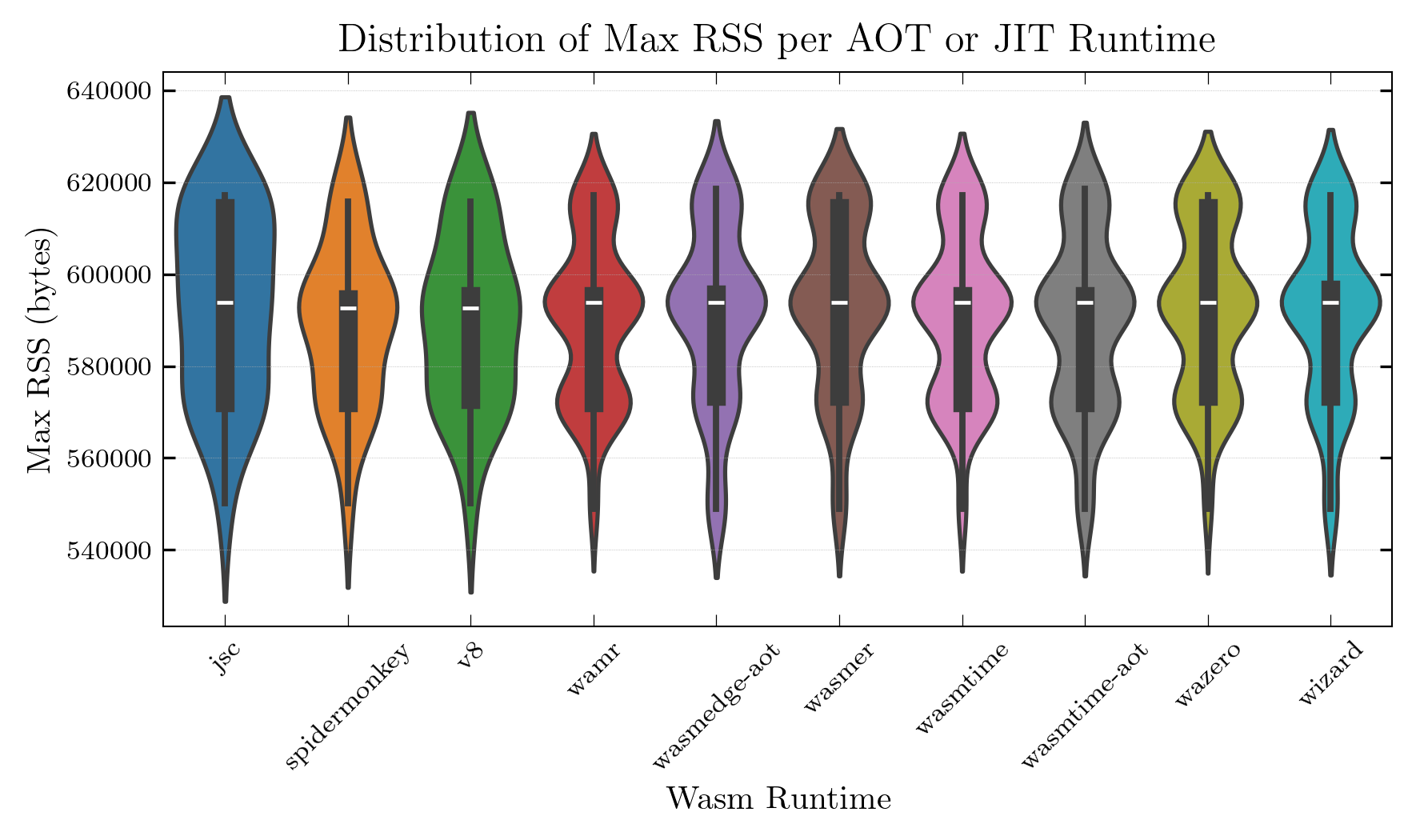}
    \caption{Distribution of maximum resident set size (Max RSS) across AOT and JIT WebAssembly engines.}
    \Description{Violin plot showing the distribution of Max RSS (in bytes) per WebAssembly engine, including jsc, spidermonkey, v8, wamr, wasmedge-aot, wasmer, wasmtime, wasmtime-aot, wazero, and wizard. The plot captures the spread and median of memory usage for each engine.}
    \label{fig:results-mean-rss}
\end{figure}

\paragraph{Runtime memory usage.}
Figure~\ref{fig:results-mean-rss} shows the distribution of maximum resident set size (Max RSS) for each WebAssembly engine. While most engines exhibit similar memory usage patterns the three browser engines (\texttt{jsc}, \texttt{spidermonkey}, and \texttt{v8}) show slightly different distribution shapes, with broader spreads and higher tails. These variations may reflect underlying differences in engine architecture and memory management strategies between embedded VMs and production-grade JavaScript engines. Overall, max RSS remains within a relatively narrow band, suggesting moderate and predictable memory footprints across the evaluated engines.

\subsection{Principal Component Analysis}
\label{sec:pca}

\paragraph{Methodology}
To study the differences between benchmark suites, we visually compare them by means of scatter plots. Rather then analyzing benchmarks on a coordinate system composed of a dimention per metric, we resort to \textit{principal component analysis (PCA)} \cite{pearson1901} to reduce data dimensionality, comparing benchmarks on the resulting coordinate system. Principal Component Analysis (PCA) performs a linear transformation of the data such that the first principal component captures the maximum possible variance. Each subsequent component then captures the highest remaining variance, subject to the constraint of being orthogonal to all preceding components. Since variance can be interpreted as a proxy for information, PCA enables dimensionality reduction by selecting only the leading components—those that retain the most variance—while attempting to preserve as much of the original information as possible.

We make use of PCA in the following way. Let \( X \) be the matrix containing the observed metrics. Each entry \( x_{ij} \in X \) (where \( i \in [1, N] \) and \( j \in [1, K] \); with \( N \) representing the number of benchmarks and \( K \) the number of metrics collected) corresponds to the normalized value of metric \( j \) for benchmark \( i \). Each metric vector $X_j$ is normalized to have a mean of zero and a standard deviation of one, resulting in a transformed vector $Y_j = \frac{X_j - \bar{x}_j}{s_j} \in Y$, where $\bar{x}_j$ and $s_j$ denote the mean and standard deviation of $X_j$, respectively. Following standard methodology, Principal Component Analysis (PCA) is performed on the normalized matrix $Y$ rather than the original matrix $X$. PCA generates a new matrix $S = YL$, where each row vector in $S$ represents the projection (or \textit{score}) of the corresponding row in $Y$ onto the principal component basis. The element $l_{ij} \in L$ (with $l_{ij} \in [-1, 1]$ and $i, j \in [1, K]$) indicates the \textit{loading} of the $i$-th metric on the $j$-th principal component (denoted as $\text{PC}_j$). The magnitude of a loading reflects the strength of association between the metric and the principal component, while the sign indicates the direction of the correlation.

\begin{table}
    \caption{Loadings of metrics on the first four principal components (PCs), sorted by absolute value (descending order).}
    \centering
    \tiny
    \begin{tabular}{|l r|l r|l r|l r|}
    \toprule
    \multicolumn{2}{|c|}{\textbf{PC1}} & \multicolumn{2}{c|}{\textbf{PC2}} & \multicolumn{2}{c|}{\textbf{PC3}} & \multicolumn{2}{c|}{\textbf{PC4}} \\
    \textbf{Metric} & \textbf{Load}. & \textbf{Metric} & \textbf{Load}. & \textbf{Metric} & \textbf{Load}. & \textbf{Metric} & \textbf{Load}. \\
    \midrule
    reach\_90 & +0.378 & writes & +0.437 & func\_cov & +0.438 & exec\_funcs & +0.524 \\
    reach\_95 & +0.378 & exec\_inst & +0.436 & block\_cov & +0.396 & total\_funcs & +0.432 \\
    reach\_75 & +0.377 & int\_ops & +0.431 & instr\_cov & +0.393 & reach\_100 & +0.320 \\
    reach\_50 & +0.376 & reads & +0.415 & rss & +0.181 & g\_writes & +0.248 \\
    reach\_99 & +0.373 & time\_ns & +0.388 & g\_writes & +0.145 & instr\_cov & +0.206 \\
    g\_reads & +0.370 & func\_cov & +0.197 & ind\_call & +0.110 & ind\_call & +0.202 \\
    block\_cov & +0.231 & instr\_cov & +0.108 & float\_ops & -0.041 & func\_cov & +0.202 \\
    instr\_cov & +0.198 & g\_writes & +0.090 & g\_reads & -0.042 & block\_cov & +0.086 \\
    reach\_100 & +0.187 & in\_first & +0.070 & reach\_50 & -0.045 & vms & +0.058 \\
    func\_cov & +0.090 & block\_cov & +0.069 & in\_first & -0.052 & reach\_99 & -0.014 \\
    exec\_funcs & +0.076 & ind\_call & +0.044 & reach\_75 & -0.055 & time\_ns & -0.018 \\
    rss & +0.039 & vms & +0.028 & time\_ns & -0.062 & int\_ops & -0.025 \\
    total\_funcs & +0.037 & float\_ops & -0.021 & reach\_90 & -0.071 & exec\_inst & -0.035 \\
    time\_ns & +0.032 & g\_reads & -0.026 & reach\_95 & -0.086 & writes & -0.052 \\
    g\_writes & +0.029 & func\_covs & -0.029 & int\_ops & -0.105 & rss & -0.059 \\
    int\_ops & +0.014 & reach\_50 & -0.034 & exec\_inst & -0.114 & reads & -0.067 \\
    exec\_inst & +0.012 & reach\_100 & -0.034 & writes & -0.127 & reach\_95 & -0.074 \\
    writes & +0.009 & reach\_75 & -0.037 & reads & -0.133 & float\_ops & -0.092 \\
    vms & +0.009 & reach\_90 & -0.039 & reach\_99 & -0.118 & reach\_90 & -0.102 \\
    ind\_call & +0.006 & reach\_95 & -0.042 & reach\_100 & -0.208 & reach\_75 & -0.128 \\
    reads & +0.005 & reach\_99 & -0.048 & vms & -0.212 & g\_reads & -0.129 \\
    in\_first & +0.005 & total\_funcs & -0.071 & func\_covs & -0.261 & reach\_50 & -0.143 \\
    float\_ops & -0.029 & rss & -0.091 & total\_funcs & -0.372 & cycles & -0.259 \\
    cycles & -0.083 & cycles & -0.121 & cycles & -0.191 & in\_first & -0.287 \\
    \bottomrule
    \end{tabular}
    \label{tab:pca-loadings}
\end{table}

\begin{figure}
    \centering
    \includegraphics[width=\linewidth]{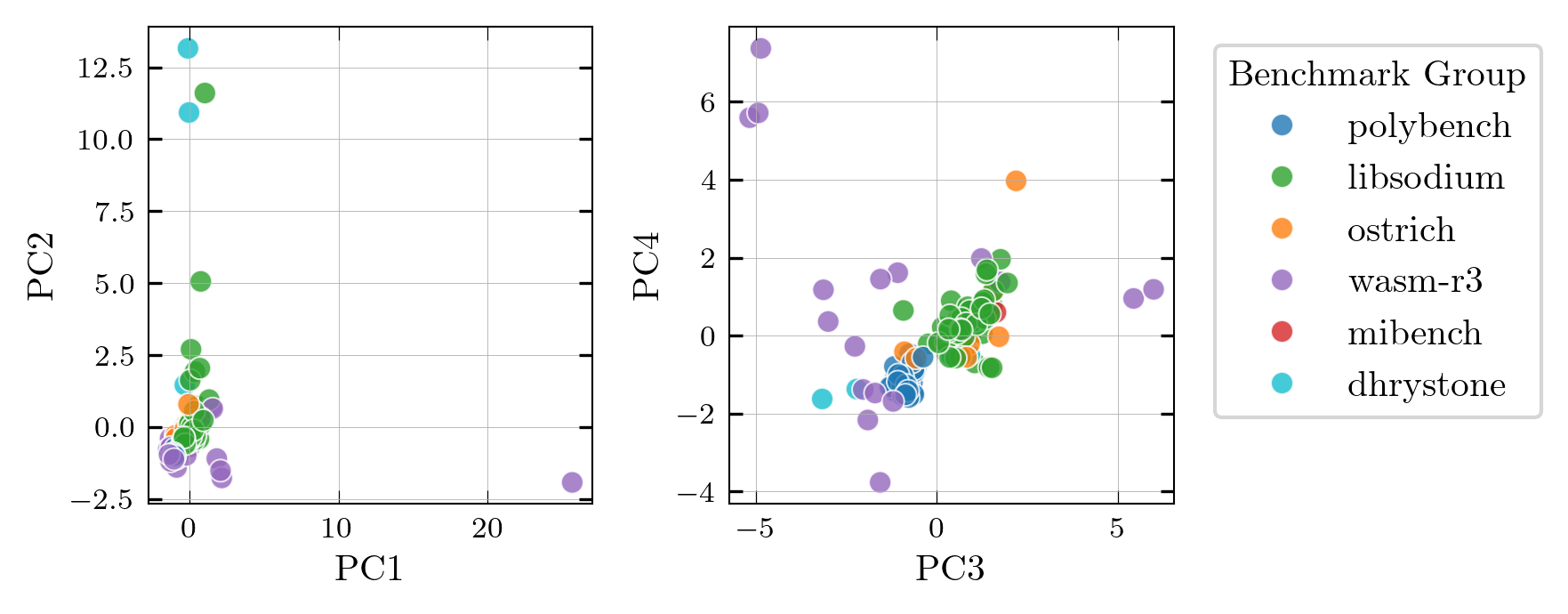}
    \Description{Two scatter plots that each show two PCs of the PCA analysis. In the first plot benchmarks are clustered towards the center, in the second one they are more spread out.}
    \caption{Scatter plots of benchmark scores over the first four principal components (PCs)}
    \label{fig:pca-scatter-plots}
\end{figure}

\paragraph{Analysis}
Table~\ref{tab:pca-loadings} shows the loading of each metric on the first four PCs, ranked by their absolute values in descending order. In Figure~\ref{fig:pca-scatter-plots} we plot the scores of the considered benchmarks against all the four PCs. The first four components account for ~60\% of the variance present in the data.

As shown in Table~\ref{tab:pca-loadings}, PC1 is primarly influenced by reach metrics and global reads. PC2 correlates strongly with dynamic execution metrics such as the number of int operations, memory reads and writes, and function coverage. Figure~\ref{fig:pca-scatter-plots} shows that benchmarks are generally tightly clustered near the origin. libsodium and dhrystone are positioned higher along PC2, reflecting their emphasis on intensive integer operations and memory activity. Interestingly, wasm-r3 exhibits a narrow spread along PC1 and presents a significant outlier characterized by high instruction reach diversity and comparatively lower execution intensity.

Figure~\ref{fig:pca-scatter-plots} reveals different structural aspects of the programs. PC3 is positively correlated with coverage metrics (block, instructions and function coverage) and negatively with memory accesses and integer operations. PC4 correlates most strongly with the number of executed and total functions, and to a lesser extent with instruction coverage. Wasm-r3 benchmarks dominate the upper region of PC4, indicating broad function execution and high coverage, while libsodium appears skewed toward negative PC3, denoting workloads that emphasize low-level operations over structural code coverage. The remaining suites (polybench, mibench, and ostrich) remain near the center across both axes, suggesting more homogeneous and less structurally diverse workloads, as we discussed previously.

\subsection{Insights and Implications}
\label{sec:analysis-implications}

\paragraph{Wasm execution behavior trends.}
Results paint a multifaceted picture of WebAssembly (Wasm) engine behavior, characterized by sharp contrasts across benchmark groups in terms of coverage, execution patterns, and memory usage. A recurring theme is the concentration of execution within a small subset of functions, especially in traditional compute benchmarks, versus the broader function utilization in real-world replay traces like \texttt{wasm-r3}. While suites like \texttt{dhrystone} and \texttt{polybench} exhibit tight loops and deterministic hot paths, \texttt{wasm-r3} reveals the noisy, high-variance execution profile of interactive web applications. Memory and global variable access patterns similarly vary: some groups (e.g., \texttt{libsodium}, \texttt{dhrystone}) exhibit intense memory traffic and minimal global use, while others (\texttt{wasm-r3}) show elevated global access.

\paragraph{Benchmark design considerations.}
The disparities in dynamic behavior across benchmark groups underscore the importance of intentional benchmark selection when evaluating WebAssembly performance and performing analysis. Suites such as \texttt{polybench} and \texttt{ostrich} are ideal for stress-testing numerical kernels or floating-point arithmetic pipelines but are poor stand-ins for workloads involving control flow diversity, global state manipulation, or real-world interaction. Conversely, replay-based benchmarks like \texttt{wasm-r3} capture realistic traces with richer control and memory profiles, albeit at the cost of higher variance and complexity. Therefore, future benchmarking efforts should include a balanced portfolio: synthetic and deterministic benchmarks for micro-architectural insights, and real-world traces for assessing engine systems, JIT heuristics, and end-user performance.

\section{Related Work}
%DONE: check DaCapo benchmark if we cited it. new paper in 2025 for da capo, cite it.

\paragraph{WebAssembly} WebAssembly (Wasm) was first announced in 2015 and has been supported by all major browser engines—Chrome, Edge, Firefox, and Safari—since 2017. Its core design and type system were formalized by Haas et al. \cite{haas2017bringing}, while Watt provided a mechanized and machine-checked proof of its soundness \cite{watt2018}. The WebAssembly standard has since evolved through various language extensions \cite{disselkoen2019, multi-memory, reference-types}, and formal efforts have further solidified its specification.

\paragraph{Dynamic Analysis of WebAssembly} Dynamic analysis techniques are increasingly being adapted to the WebAssembly ecosystem. Several early works introduced runtime analyses for Wasm, including two taint-tracking systems \cite{fu2018taintassembly, szanto2018taint} and a cryptomining detector \cite{wang2018}. These tools differ substantially in implementation complexity and approach: from modifying the V8 JavaScript engine \cite{fu2018taintassembly}, to developing a new Wasm interpreter in JavaScript \cite{szanto2018taint}, to employing binary instrumentation techniques \cite{wang2018}.

Wasabi \cite{lehmann2019} presents a general-purpose framework for dynamic analysis of WebAssembly, aiming to lower the barrier to implementing such analyses. Built on top of binary instrumentation, Wasabi enables developers to implement a range of analyses—including taint tracking and control-flow profiling—with significantly reduced effort compared to engine-level modifications or custom VMs.

More recently, Titzer et al. \cite{titzer2024}  introduced Wizard, a novel research-oriented WebAssembly engine that supports non-intrusive, dynamic instrumentation. Unlike previous approaches that required modifying the engine or performing static instrumentation, Wizard provides a flexible infrastructure for injecting and removing bytecode-level probes at runtime. Its design enables composable, high-level analyses to be constructed from low-level primitives while leveraging the engine's JIT compiler, interpreter, and deoptimization mechanisms. Wizard offers consistency guarantees and performance optimizations that make it suitable not only for research but potentially for production environments, especially when instrumentation is disabled. This contribution marks a significant step forward in the expressiveness and practicality of dynamic instrumentation for WebAssembly.

Recent frameworks such as Wastrumentation \cite{munsters2025} and Whamm \cite{gilbert2025} extend this line of work by providing portable, high-performance instrumentation infrastructures—Wastrumentation enabling intercessive, source-level analyses, and Whamm offering a declarative DSL for engine-integrated instrumentation.

Together, these systems—Wasabi and Wizard—demonstrate that dynamic analysis of WebAssembly is both feasible and evolving rapidly. Each provides a complementary approach to the growing need for better runtime observability in this increasingly popular compilation target.

\paragraph{Dynamic Analysis in General} Dynamic analysis has long been used as a complementary approach to static analysis \cite{ball1999, ernst2004}. Over the years, a variety of dynamic analyses have been proposed, including dynamic slicing \cite{agrawal1990}, taint analysis for x86 binaries \cite{newsome2018} and Android apps \cite{enck2014}, as well as analyses to uncover concurrency issues \cite{lu2006, park2009, burckhardt2010}, memory errors \cite{chilimbi2006}, value tracking \cite{bond2007}, and performance bottlenecks \cite{yu2014}. The increasing popularity of WebAssembly as a compilation target for various languages and applications suggests a growing need for dynamic analysis tools tailored to its unique execution model. Complementary research on computation-intensive loop kernels \cite{hashimoto2017} highlights the broader need to characterize performance-critical code patterns, informing both dynamic and static optimization strategies.

\paragraph{Studies of WebAssembly} Several studies have examined the real-world usage and performance characteristics of WebAssembly. Musch et al. \cite{musch2019} performed one of the first large-scale investigations by systematically collecting WebAssembly binaries from websites. Their findings revealed that cryptomining was a dominant use case at the time. More recent work, however, has challenged this notion. The WasmBench study \cite{hilbig2021} builds on and significantly extends Musch et al.’s methodology by collecting 58× more binaries from a broader range of sources. This larger dataset enables a more comprehensive analysis, demonstrating that cryptomining is no longer the predominant application and that a wider variety of domains—including gaming, multimedia, and utility libraries—have adopted WebAssembly. 

\paragraph{WebAssembly Benchmarks} A recurring challenge in evaluating WebAssembly performance is the absence of a comprehensive, standardized benchmark suite tailored to the characteristics and use cases of real-world WebAssembly binaries. Much of the existing work relies on general-purpose or domain-specific benchmark suites originally designed for other platforms. These include PolyBenchC, SciMark, and Ostrich \cite{haas2017bringing, lehmann2019, herrera2018, khan2014}, which focus primarily on numerical and scientific computations, as well as SPEC CPU \cite{jangda2019, lehmann2019}, which targets complex C/C++ programs typically not representative of workloads compiled to WebAssembly in practice. Additionally, some studies rely on small, hand-curated sets of applications \cite{lehmann2019}, which often fail to capture the diversity and scale of WebAssembly usage in the wild. The DaCapo benchmark suite \cite{blackburn2006} has long served as a model for principled benchmarking methodology in managed runtimes, influencing subsequent benchmark design across languages and environments, including ongoing work in WebAssembly.

This lack of representative and reproducible benchmarking infrastructure poses a barrier to systematic performance evaluation of WebAssembly engines and toolchains. Without a standardized suite or automation framework, benchmarking efforts are often ad hoc, difficult to reproduce, and limited in scope. This gap highlights the need for a dedicated tool that streamlines the benchmarking process and facilitates consistent, large-scale performance evaluation of WebAssembly binaries.

\section{Conclusion}

In this paper, we introduced Wasure, a command-line toolkit for benchmarking WebAssembly engines. Wasure was designed to be modular, extensible, and easy to use, providing a streamlined workflow for evaluating runtime performance, compatibility, and feature support across a wide range of WebAssembly engines.

We described the design goals, architecture, and core modules of the system, highlighting how it enables reproducible and customizable experiments. Wasure supports fine-grained engine management, flexible benchmark definitions, and multiple output formats for downstream analysis.

To complement performance evaluation, we conducted a dynamic analysis of the benchmark suite included with Wasure. Using external instrumentation via the Wizard engine, we profiled various runtime characteristics such as code coverage, instruction hotness, memory behavior, and control flow diversity. These insights revealed distinct behavioral patterns across benchmark groups, ranging from tight, deterministic kernels to broad, replay-based traces. This highlightes the importance of benchmark diversity in WebAssembly evaluation.

Wasure aims to serve professionals and researchers by lowering the barrier to comparative performance analysis and facilitating more systematic benchmarking efforts in the WebAssembly ecosystem.

%%
%% The acknowledgments section is defined using the "acks" environment
%% (and NOT an unnumbered section). This ensures the proper
%% identification of the section in the article metadata, and the
%% consistent spelling of the heading.

%%
%% The next two lines define the bibliography style to be used, and
%% the bibliography file.
\bibliographystyle{ACM-Reference-Format}
\bibliography{biblio}

%%
%% If your work has an appendix, this is the place to put it.
\appendix

\end{document}